\documentclass[12pt]{article}
\usepackage{latexsym}
\usepackage{amssymb,amsmath,bm}
\usepackage{graphicx}
\usepackage{color}
\usepackage[usenames, dvipsnames]{xcolor}
\usepackage{tikz}
\usepackage[colorlinks=true]{hyperref}

\usepackage[top=1.25in,bottom=1.25in,right=0.85in,left=0.85in]{geometry}

\usepackage[numbers,merge,sort&compress]{natbib}

\newcommand{\refl}[1]{(\ref{#1})}
\newcommand{\beq}{\begin{equation}}                                             
\newcommand{\eeq}{\end{equation}}     
\newcommand{\eeql}[1]{\label{#1}\eeq}

\newcommand{\x}{\ensuremath{\times}}
\newcommand{\ra}{\ensuremath{\rightarrow}}
\newcommand{\zpr}{\ensuremath{Z'}}

\newcommand{\uprm}{\ensuremath{U(1)'}}
                                          
\newcommand{\oh}{\ensuremath{\frac{1}{2}}}

\newcommand{\stto}{\ensuremath{SU(2)_{L} \x SU(2)_{R} \x U(1)}}                                           
\newcommand{\sto}{\ensuremath{SU(2) \x U(1)}}                                       
\newcommand{\st}{\ensuremath{SU(2)}}                                                                                       
\newcommand{\sth}{\ensuremath{SU(3)}} 
\newcommand{\stf}{\ensuremath{SU(5)}}     

\newcommand{\vect}[1]{\ensuremath{
\left( \begin{array}{c} #1    \end{array} \right) }}

\newcommand{\veva}[1]{\ensuremath{\langle#1\rangle}}

\newcommand{\ov}{\overline}

\newcommand{\cM}{\mathcal{M}}
\newcommand{\cN}{\mathcal{N}}
\newcommand{\bZ}{\mathbb{Z}}
\newcommand{\bF}{\mathbb{F}}
\newcommand{\bP}{\mathbb{P}}
\newcommand{\bC}{\mathbb{C}}         

\def\ge{E}
\def\gso{SO}
\def\gsu{SU}
\def\gsp{Sp}
\def\gf{F}

\def\gg{G}

\newcommand{\drawsquare}[2]{\hbox{%
\rule{#2pt}{#1pt}\hskip-#2pt
\rule{#1pt}{#2pt}\hskip-#1pt
\rule[#1pt]{#1pt}{#2pt}}\rule[#1pt]{#2pt}{#2pt}\hskip-#2pt
\rule{#2pt}{#1pt}}
\newcommand{\fund}{\raisebox{-.5pt}{\drawsquare{6.5}{0.4}}}
\newcommand{\Ysymm}{\raisebox{-.5pt}{\drawsquare{6.5}{0.4}}\hskip-0.4pt%
\raisebox{-.5pt}{\drawsquare{6.5}{0.4}}}
\newcommand{\Yasymm}{\raisebox{-3.5pt}{\drawsquare{6.5}{0.4}}\hskip-6.9pt%
\raisebox{3pt}{\drawsquare{6.5}{0.4}}}
\newcommand{\antifund}{\overline{\fund}}

\newcommand{\jh}[1]{{#1}}

\begin{document}

\vspace*{.8cm}
\begin{center} {\LARGE\jh{TASI Lectures on} Remnants from the String Landscape}

\vspace*{1.5cm}
James Halverson$^{1}$ and Paul Langacker$^{2,3}$
\vspace*{1.0cm}

$^1${\it Departments of
  Physics, Northeastern University, \\ Boston, MA 02115} \\ 
{\tt jhh@neu.edu} \\ \vspace{.5cm}

$^2${\it School of Natural Science, Institute for Advanced Study, \\ Princeton, NJ 08540}
 \\ \vspace{.2cm}
  $^3${\it Princeton University, Princeton, NJ 08544}\\
{\tt pgl@ias.edu}

\vspace*{0.8cm}
\end{center}
\vspace*{0cm}

Superstring theories are very promising theoretically, but the enormous landscape of string vacua and the
(likely) very large underlying string scale imply that they may never be tested directly. Nevertheless,
concrete constructions consistent with the observed world frequently lead to observable remnants, i.e., new particles or features that are apparently accidental consequences of the ultraviolet
theory and that are typically not motivated by specific shortcomings of the standard models
of particle physics or cosmology. For example, moduli, axions, large extended gauge sectors, additional $Z'$ gauge bosons, extended Higgs/Higgsino sectors, and quasi-chiral exotics are
extremely common. They motivate alternative cosmological paradigms and could lead to
observable signatures at the LHC.
Similar features can emerge in other standard model extensions, but in the stringy case they are more likely to occur in isolation and not as part of a more complete TeV-scale structure. Conversely, some common aspects of the infinite ``landscape'' of field theories, such as large representations, are expected to be very rare in the string landscape, and observation
of features definitively in the swampland could lead to falsification. In this article, common stringy remnants and their phenomenology are surveyed, and  implications for indirectly supporting or  casting doubt on string theory are discussed.

\clearpage
\tableofcontents

\newpage

\section{Introduction}

Superstring theory\footnote{For a review, see, e.g.,~\cite{Ibanez:1428137}.} yields a consistent and  finite unification of quantum theory, gravity, and  other interactions, at least at the perturbative level, and is therefore a promising candidate for an ultimate unified theory. However, it is not clear whether Nature  actually takes advantage of string theory, or whether there is any way to confirm or falsify it.\footnote{For general discussions of some possibilities, 
see~\cite{Langacker:2003xa,*Barger:2007ay,Nath:2010zj,Gato-Rivera:2014afa,Acharya:2012tw,*Kane:2016cau,*Kane:2011kj,Acharya:1744041,Polchinski:2015pzt,*Polchinski:2016xto,Quevedo:2016tbh,LangackerPUP,Giudice:2017pzm}.} 
Part of the problem is that the fundamental string scale is most likely much larger than can ever be directly probed experimentally.\footnote{This problem is likely to be shared by most alternative theories of quantum gravity.} Moreover, there are actually five superstring theories,
each of which is consistent in 10 space-time dimensions.
These are connected to  one another by a number of dualities, and may all
be different limits of a strongly coupled M-theory (which also contains 11-dimensional supergravity) that is only partially understood. 

Further complications arise since there appears to be an enormous landscape of possible string vacua, each involving different low energy physics, with no known principle for the dynamical selection of vacua similar to ours. 
This situation is not unique to string theory: landscapes of vacua or metastable states
naturally exist elsewhere in physics such as protein folding \cite{doi:10.1021/j100356a007,*PROT:PROT340210302,*doi:10.1146/annurev.physchem.48.1.545} and spin glasses (e.g., ~\cite{Denef:2011ee}). They
also exist elsewhere in
high energy physics in quantum field theories with multiple scalar fields, including
many extensions of the standard model, or 
even within the standard model itself 
if one allows one or more of the space dimensions to be compact~\cite{ArkaniHamed:2007gg}.

The situation can be argued to be even worse in the conventional field theory framework. Even putting aside the question of quantum gravity, restricting to four-dimensional renormalizable theories, and ignoring possible compactifications, there is apparently an infinite landscape of possible field theories, characterized by different particles, interactions, symmetries, and parameters. There is again no known selection principle  to choose between them.

Furthermore,
  it has been argued that the landscape of string vacua may be surrounded by a much larger ``swampland'' of apparently-consistent effective field theories that are in fact inconsistent with quantum gravity or string theory~\cite{Vafa:2005ui,*Ooguri:2006in}.\footnote{
Examples include theories with an exact continuous global symmetry~\cite{Banks:1988yz,*Witten:2000dt,*Burgess:2008ri,*Banks:2010zn,*Witten:2017hdv} or in which a gauge symmetry is weaker than gravity (the ``weak gravity conjecture'')~\cite{ArkaniHamed:2006dz}; see \cite{Rudelius:2014wla,Rudelius:2015xta,Brown:2015iha,Montero:2015ofa,Bachlechner:2015qja,Heidenreich:2016aqi} for recent
weak gravity conjecture studies.} 
From this perspective, the low energy limit of string vacua yields a proper subset of the possible field theories,
due to some additional consistency conditions and constraints on possible  (or at least likely) representations and groups.
The observation of new physics in violation of these constraints would go far towards falsifying the existence of an underlying string theory. 

Conversely, string vacua that contain the standard model (SM) or MSSM and that are qualitatively consistent with what we know may contain apparently unmotivated remnant particles or features, with some types occurring frequently in the string landscape. 
Observation of some of these characteristic remnants, plus the enormous theoretical advantage of a finite quantum gravity, would lend considerable support to the notion of string theory.

String remnants are in sharp contrast with
most bottom-up and other non-stringy investigations, where new physics is motivated by attempting to solve problems of the standard model, or by notions of minimality,  elegance,  or naturalness.\footnote{There, some form of uniqueness is often implicitly assumed, but in practice that is mostly a function of the gauge groups, representations, and other symmetries that are  invoked.
As already emphasized there is no known selection principle or consistency condition leading to a unique field theory. Furthermore, the standard model itself is very complicated, most viable extensions do not appear to be very minimal, and the notion of elegance is rather subjective. For further discussion, see, e.g.,~\cite{LangackerPUP,Giudice:2017pzm}.} String remnants, on the other hand, need not solve any problems, might not appear to be minimal or elegant, and might or might not be consistent with notions of naturalness. Instead,
they are motivated simply by their frequent occurrence in string theory, for
example in surveys of semi-realistic vacua containing the standard model or MSSM.
Compactifications involving any or certain classes of remnants are often discarded 
in many papers that search for the MSSM (and, possibly, common bottom-up extensions),
despite the fact that they are often experimentally allowed.\footnote{Examples of papers that focus on general surveys of the landscape 
include~\cite{Denef:2004ze,*Denef:2004cf,Kumar:2004pv,DeWolfe:2004ns,Blumenhagen:2004xx,*Gmeiner:2005vz,Gmeiner:2006vb,Dienes:2004pi,ArkaniHamed:2005yv,DeWolfe:2005uu,Kumar:2006tn,Dienes:2006ut,Dienes:2007ms,Anastasopoulos:2006da,Shelton:2006fd,Gmeiner:2006qw,*Gmeiner:2007zz,Douglas:2006es,Acharya:2006zw,Denef:2007pq,Blumenhagen:2006ci,AbdusSalam:2007pm,Balasubramanian:2008tz,Gabella:2008id,Cvetic:2011iq,Morrison:2012np,*Taylor:2015xtz,Schellekens:2013bpa,Schellekens:2015cia,*Schellekens:2015zua,Nibbelink:2013lua,Braun:2014lwp,Halverson:2014tya,Ecker:2014hma,Watari:2015ysa,Halverson:2017ffz,Cvetic:2017epq}. Studies that are more focused on obtaining the MSSM or grand unified theories,  often with remnants, include~\cite{Faraggi:1989ka,Faraggi:1992fa,*Faraggi:1997dc,*Faraggi:2006qa,*Faraggi:2008wg,*Faraggi:2017cnh,Chaudhuri:1994cd,*Chaudhuri:1995ve,Cleaver:1997jb,*Cleaver:1998im,Cleaver:1998gc,*Cleaver:1998sm,Donagi:1999ez,Giedt:2000bi,Ibanez:2001nd,Cremades:2002dh,Cvetic:2001tj,*Cvetic:2001nr,Cvetic:2002qa,Antoniadis:2002qm,Dijkstra:2004cc,Kobayashi:2004ud,*Kobayashi:2004ya,Bouchard:2005ag,Verlinde:2005jr,Cvetic:2005bn,Blumenhagen:2005mu,Braun:2005ux,*Braun:2005nv,Buchmuller:2005jr,*Buchmuller:2006ik,*Lebedev:2006tr,Lebedev:2007hv,*Nilles:2008gq,*Lebedev:2008un,Kim:2007mt,Chen:2007px,Blumenhagen:2008zz,Beasley:2008dc,*Beasley:2008kw,Donagi:2008ca,*Donagi:2008kj,Conlon:2008wa,Marsano:2009gv,Heckman:2010bq,Raby:2011jt,Anderson:2011ns,*Anderson:2012yf,*Anderson:2013xka,Quevedo:2014xia,Honecker:2016gyz}.}

 Our attitude, on the other hand, is that such remnants could well exist, which would be a clue that there is an underlying string theory. 
Of course, such things as extended gauge, scalar, and fermion sectors are not unique to string vacua, and many possible standard model extensions could emerge either as remnants or from more conventional motivations.
The best that we can do here is to emphasize (given current understanding) that some
possible standard model extensions occur very frequently in the string landscape, while
others are absent or very rare, and to contrast with bottom-up approaches.

For example, in this paper we are mainly concerned with those vacua involving a perturbative connection between the low energy and string scales, in contrast to
the many bottom-up models that
involve some form of strong coupling or compositeness at the (multi-)TeV scale, often with a fairly large number of interactions and elementary or composite particles. 

  By a perturbative connection\footnote{The restriction to a perturbative extension is motivated by  practicality. If we were to observe a strong coupling extension of the standard model it would be extremely difficult to penetrate through the details well enough to connect to an underlying string theory, even though such sectors may arise in the landscape. Of course, other sectors, such as those that could be associated with supersymmetry breaking or dark matter, could be strongly interacting.} between the low energy
 and string scales, we mean that the standard model and cosmological degrees of freedom are essentially the ones that emerge directly from the string compactification, rather than as composite objects associated with some new strong interactions. Unlike most conventionally-motivated bottom-up models, these constructions frequently include remnants that appear ad hoc and isolated, not part of a more complete low-energy theory, and not necessarily associated with a standard model problem.

In this paper we consider the common types of remnants that are  likely to be found in semi-realistic  string vacua with a perturbative standard model sector, and some of their phenomenological consequences. We will also consider cosmological implications
of moduli and axions, and we are mainly but not exclusively concerned with the case of a  large string scale. 
 By semi-realistic we mean vacua that are not grossly inconsistent with observation, and in cases
 where particle (rather than cosmological) sectors are being considered we mean that
 the vacua contain at least the standard model or MSSM spectrum. We do not, however, insist that such details as the fermion masses and mixings  are correctly predicted. There are few, if any, known examples of constructions that successfully predict such detailed features, due in
 part to the difficulty of moduli stabilization.
 Moreover, some or all Yukawa couplings may well be essentially random features of the particular string vacuum, and
  it is not clear whether these would be strongly correlated with remnants. 
Even if string theory is correct it is unlikely that anyone will ever find the exact vacuum corresponding to our world. We therefore do not attempt to give fully realistic examples, but instead  focus on general ideas and features suggestive of string theory. 
 
Finally, if there really is an exponentially large number of string vacua\footnote{As opposed
to an exponentially large number of ``compactification possibilities'' that may give rise to
a much smaller number of vacua. It is certainly true that large numbers of possibilities exist, but concretely
demonstrating the existence of metastable vacua has thus far been limited to regimes of control with low
numbers of moduli.}, and assuming  something like eternal inflation,  then one must seriously consider the possibility of  environmental selection (e.g., as an explanation of apparent fine-tunings and fortuitous accidents) and the multiverse.\footnote{Recent general discussions include~\cite{Weinberg:2005fh,Guth:2007ng,Schellekens:2015cia,Linde:2015edk,Donoghue:2016tjk}.} We will briefly mention such notions, 
 but the thrust of this paper is instead towards apparently ad hoc extensions of the standard model
 and the possibility that they could provide evidence for string theory. These
 remnants are likely concrete lessons from string theory regardless
 of the extent to which anthropic selection
 plays a role in landscape.

There is also the practical question of how to study such large datasets, which
is discussed in section \ref{sec:bigdata}.

 \vspace{.5cm}
 
The plan of the paper is as follows. The main body is  a non-technical description of some of the possible types of string theory remnants and their phenomenological implications. In particular, Section~\ref{stringtheories}
contains a qualitative description of the various classes of string constructions, such as heterotic, Types IIA, IIB, and F-theory. Section~\ref{remnants} discusses the major classes of possible string remnants and their phenomenology.
Included are moduli and axions; extended gauge sectors, especially $U(1)'$ s; extended particle sectors, including Higgs, Higgsinos, quasi-chiral exotics, the absence of large representations, and (quasi-)hidden sectors; implications for couplings and hierarchies, such as leptoquarks, diquarks, family nonuniversality and flavor change, Yukawa interactions/hierarchies, and nonstandard mechanisms for neutrino mass; implications for grand unification and gauge unifications,
a low string scale, and environmental selection; and more exotic possibilities.
Section~\ref{discussion} is a brief discussion and summary.
The appendices contain  theoretical details of some of the possibilities, including dark sectors in F-theory, axions, moduli, 
and stringy consistency conditions with implications for exotics. Most of the content in the
appendices was presented by J.H. at TASI 2017.

\section{String Theory}\label{stringtheories}
\subsection{General Features}
In this section we provide a brief qualitative discussion of string theory and its compactifications. 
For each major topic introduced we cite  critical original literature  or reviews to aid 
the reader in further study.

Superstring theory is a quantum theory of gravity that is a leading candidate for
unification. It naturally exhibits a massless
graviton in its spectrum, and the amplitudes for two to two graviton scattering are finite in perturbation theory. 
Its low energy
effective action contains the Einstein-Hilbert action crucial to general relativity, as well as other fields.  
Superstring theory exists in ten dimensions, and there are five consistent possibilities that 
are known as heterotic $E_8\times E_8$, heterotic
$SO(32)$, type I, type IIA, and type IIB. Non-abelian gauge sectors may arise in each of these theories,
due either to fundamental excitations of closed strings propagating in ten dimensions, or to open strings (or their
generalizations) that end on solitonic objects collectively referred to as branes, or D-branes in more specialized
cases.

These theories are related to one another by a number of dualities.
For example, T-duality exchanges two string theories compactified on circles of radius $R$ and $1/R$;
type IIA and type IIB are T-dual to one another, as are heterotic $E_8\times E_8$ and heterotic $SO(32)$.
In the second superstring revolution of the mid-1990's dualities between weakly coupled and strongly coupled theories
were also demonstrated, known as S-dualities. The type I and heterotic $SO(32)$ theory are S-dual to one another,
and the type IIB theory is self-dual under S-duality. More generally, the type IIB theory allows for the string
coupling to vary over the extra dimensions, 
a possibility known as F-theory \cite{Vafa:1996xn}. 
The weakly coupled type IIA and heterotic $E_8\times E_8$ theories are weakly coupled limits of an eleven-dimensional
theory known as M-theory \cite{Witten:1995ex}, which is a quantum ultraviolet completion of eleven-dimensional
supergravity. Via this web of dualities, the different superstring theories in ten dimensions are different limits
of one theory. For  canonical textbooks on superstring theory and each duality, see ~\cite{Ibanez:1428137,Green:110204,Becker:1003112,polchinski1998string,*polchinski1998string2,Zwiebach:1166312,Blumenhagen:1493265}.

\vspace{.5cm}
The universe that we observe around us has four apparent dimensions. However, as shown by Kaluza and Klein in 
the 1920's, a higher-dimensional theory may appear to have fewer dimensions at low energies if the extra dimensions
are small and compact. This general result may be applied to superstring theory: 
if some of its ten dimensions are along a compact $d$-manifold, the low energy effective physics would be 
$(10-d)$-dimensional, in which case the choice $d=6$ recovers a four-dimensional universe. In many cases this
low energy effective physics is well approximated by a low energy effective Lagrangian. One of the most studied
possibilities is to arrive at four-dimensional physics with $\cN=1$ supersymmetry\footnote{Chiral fermions arise naturally from theories with $\cN=0$ and $\cN=1$ supersymmetry, 
which is part of the reason for the focus on $\cN=1$ supersymmetry.}
via compactification on
a Calabi-Yau threefold (CY3) \cite{Candelas:1985en}, which has complex dimension three or real dimension six.

Given the framework of compactification, it is natural to consider the compactification of each of the superstring
theories, of F-theory, or of M-theory. This has been the subject of a research program over the last thirty
years that is sometimes referred to as string phenomenology, though often the research is formal in nature
and is not immediately concerned with particle physics or cosmology. 
Given the choice of one of the higher-dimensional theories there are often many four-dimensional
solutions, which are often referred to as the string landscape. 

 \vspace{.5cm}
 The remainder of this section contains a more technical treatment of some  well-studied possibilities
that preserve $\cN=1$ supersymmetry at leading order. The reader who is uninterested in these details may go directly to the 
phenomenological discussion in Section~\ref{remnants}.

\begin{itemize}
\item \textbf{Heterotic on CY3.} Compactification of the heterotic string~\cite{Gross:1984dd,*Gross:1985fr,*Gross:1985rr} on a Calabi-Yau threefold $X$ is 
	one of the first compactification classes that was studied. In addition to the choice of $X$, one must
	also make the choice of a gauge field background, as encoded in a holomorphic vector bundle $V$ on $X$.
	The four-dimensional gauge group $G$ is the commutant of the structure group of $V$ inside $E_8$. There
	are two central classes of models:
	\begin{itemize}
	\item \emph{Standard Embedding}. In this case $V=TX$, the tangent bundle to $X$, in which case the structure group
	is $SU(3)$ and $G=E_6\times E_8$. The $E_8$ gauge sector is a hidden sector and the $E_6$ factor may therefore
	give rise to grand unification, where the number of chiral particle generations is $\chi(X)/2$, where $\chi(X)$ is the
	Euler character of $X$.
	\item \emph{Non-standard Embedding}. In this case $V \neq TX$, and there are a number of constructions for
	$V$, such as spectral covers or monads. The structure group, and therefore the four-dimensional gauge group $G$,
	varies depending on topological choices. $SU(5)$ or $SO(10)$ grand unification may arise from the choice of
	$SU(5)$ or $SU(4)$ structure group, respectively.
	\end{itemize}
	In both cases the low energy effective action contains scalar fields that are massless at leading order.
	The scalar fields that correspond to metric deformations of $X$ that preserve the Calabi-Yau condition are
	known as complex structure or K\" ahler moduli, and the scalar fields that correspond to holomorphic deformations
	of $V$ that preserve the Hermitian Yang-Mills equations are known as vector bundle moduli. Models with the standard
	embedding may in certain cases be specializations in the vector bundle moduli space of non-standard embedding
	models.
\item \textbf{Type II Compactifications with Orientifolds, D-branes, and Fluxes.} Compactifications of the type
	II superstring theories on a Calabi-Yau threefold $X$ preserve $\cN=2$ supersymmetry. However, D-branes and 
	negative tension objects called orientifold planes may be introduced into the compactification that break half
	of the supersymmetry, yielding a $4d$, $\cN=1$ compactification. 
	Organizing according to
	the origin of gauge sectors, subclasses of these compactifications include
	\begin{itemize}
		\item \emph{Type IIA Intersecting Branes.} Here an O6-plane is located on the fixed point locus of an
			antiholomorphic involution on $X$, and stacks of spacetime filling D6-branes exist to cancel the Ramond-Ramond
			charge of the O6-plane. Diverse gauge sectors arise from the D6-branes and matter charged under the 
			gauge groups arises from intersections of D6-brane stacks in $X$, which occur at points. 
		\item \emph{Type IIB Compactifications with D7-branes.} In these compactifications an O7-plane is located on
		the fixed point locus of a holomorphic involution on $X$ and spacetime filling D7-branes exist to
		cancel the Ramond-Ramond charge of the O7-plane. D7-branes may intersect, though they now do so along
		holomorphic curves, which have real dimension two, unlike points. Vector bundles or worldvolume fluxes
		on the branes give rise to chiral matter at their intersections, as can be understood in simple cases 
		via T-duality, which exchanges the number of intersection points of D6-brane stacks with properties of
		worldvolume fluxes on seven-branes. 
		\item \emph{Type IIB Compactifications with Magnetized D9-branes.} Here gauge sectors arise from spacetime filling
		D9-branes and D5-branes, and again appropriate Gauss laws must be satisfied to ensure the net cancellation
		of Ramond-Ramond charge in the extra dimensions. The structure of chiral matter from the D9-branes depends on the
		choice of bundle or magnetic fluxes on the D9-branes. In simple cases, these can be understood from
		two T-dualities of configurations with D7-branes.
		\item \emph{Type IIB with D3-branes at Singularities.} $X$ may have point-like singularities, such as 
		orbifold points. A D3-brane at or near such a singular point gives rise to a non-trivial quiver gauge theory,
		the structure of which is determined by the structure of the singularity. Such D3-brane sectors may also
		arise in compactifications with D7-branes. 
	\end{itemize}
For reviews of this class of models, see \cite{Blumenhagen:2005mu} and the very thorough \cite{Blumenhagen:2006ci}.
For D3-branes at singularities, see  \cite{Kennaway:2007tq,Yamazaki:2008bt}.
\jh{For generalized structures in type II compactifications that lead to $\cN=1$ vacua,
see \cite{Grana:2005sn,Grana:2006kf}.}

Additionally, background fluxes may be turned \cite{Dasgupta:1999ss,Giddings:2001yu} on in type IIB compactifications that may stabilize the moduli of $X$; \jh{see \cite{DeWolfe:2005uu,McOrist:2012yc} for studies of type IIa flux compactifications.}

The best understood scenarios are type IIB flux compactifications with three-form background fluxes, which are used
in the KKLT \cite{Kachru:2003aw} and Large Volume Scenario \cite{Balasubramanian:2005zx} (LVS) moduli stabilization schemes.
The number of choices of such fluxes depends on $X$ but is exponentially large and may give rise to a small cosmological
constant \cite{Bousso:2000xa}. Given the large number of possibilities and the associated computational complexity,
a statistical approach~\cite{Douglas:2003um} has been advocated.

\item \textbf{F-theory on elliptic CY4.}
Weakly coupled type IIB orientifold compactifications are a subset of a larger set of constructions known as F-theory,
which may have weakly coupled limits \cite{Sen:1996vd,Sen:1997gv}. However, in one of the two largest known
ensembles it was shown \cite{Halverson:2017vde} that such limits essentially never exist.

In F-theory the axiodilaton field of type IIB compactifications 
is treated as the complex structure of an auxiliary elliptic
curve that varies (holomorphically) over the compact extra dimensions $B$. Since the axiodilaton determines the string
coupling $g_s$, the latter also varies over the spacetime. The appropriate geometric
structure is a CY4 $X$ that is elliptically fibered over $B$. The structure of four-dimensional gauge theories
depends on the singular fibers of $X$, which encodes information about seven-branes stacks, and also fluxes on
the seven-branes that determine the chiral spectrum. Seven-branes in F-theory are more general than those of
weakly coupled type IIB compactifications; for example, they may contain exceptional gauge groups.

The generalized seven-brane structure gives rise to interesting possibilities for grand unification \cite{Beasley:2008dc,Donagi:2008ca} that do not exist at weak coupling, such as a perturbative top-quark Yukawa coupling. See \cite{Heckman:2010bq,Weigand:2010wm} and references therein for additional discussion of grand unification in F-theory. See 
appendix \ref{sec:dark gauge Ftheory} for a more detailed introduction to F-theory, including non-Higgsable clusters, or \cite{Weigand:2010wm,Heckman:2010bq}.

\item \textbf{M-theory on $G_2$ manifolds.}
Since M-theory is an eleven-dimensional theory, obtaining a four-dimensional compactification requires that the extra
dimensions of space have seven dimensions. To obtain a $4d$ theory with $\cN=1$ supersymmetry, the seven-dimensional
analog of a Calabi-Yau threefold is a seven-manifold $X$ with $G_2$ holonomy.

If $X$ is smooth and large, the $4d$ low energy effective theory has $\cN=1$ supersymmetry and has $b_2(X)$ massless
abelian vector multiplets, but no charged matter, and $b_3(X)$ massless chiral multiplets arising from metric deformations
that preserve the $G_2$ holonomy and the three-form $C_3$ of eleven-dimensional supergravity. If $X$ is fibered by a circle,
it becomes a type IIA compactification with O6-planes and intersecting D6-branes in the limit that the
circle becomes small.

Obtaining a non-abelian gauge sector with chiral matter therefore requires taking a limit in which $X$ becomes singular,
in which case gauge sectors may arise along real codimension four ADE singularities, with vector pairs of 
charged chiral multiplets arising from codimension six singularities and charged chiral matter arising at
codimension seven singularities. See the review \cite{Acharya:2004qe} and references therein.

Each of these constructions has its own nuances and specialized considerations, but some low energy implications arise
out of many of the constructions, which is the idea behind remnants and the subject of this review.

\end{itemize}
There are other well-studied constructions beyond those discussed here. For an introduction to those, as well as a
thorough introduction to all of the compactification classes mentioned in this section, we refer the reader to
the excellent textbook
\cite{Ibanez:1428137}. 

\subsection{Remarks on Universality, Machine Learning, and the Landscape}
\label{sec:bigdata}

 There is also the practical question of how to study datasets that are as large
 as the ones that arise in string theory, where the number of data points is
 far larger than the $O(10^{78})$ protons that exist in the visible universe.
 Answering it is of great importance for understanding remnants.

 As an
 analogy, consider large energy scales in particle physics. Successive generations
 of colliders probe successively higher energy scales, but there is a demarcation line for
 what constitutes a ``large" energy scale, which is one so large that no conceivable collider
 will ever probe it directly. For many reasons, including uncertainties in the rate of
 progress in accelerator physics and in science funding, the value of this demarcation
 line is not clear. However, physicists know values above this line when they see it:
 we will almost certainly never directly probe the GUT scale or Planck scale, or the
 string scale if it is large.

 Similarly, one good notion of what constitutes a large dataset, 
 including in the context of the string landscape, is a dataset so large that no
 conceivable computer will ever be able to hold it in memory. Since each data point
 must be at least one bit of information, the geometric ensemble of \cite{Halverson:2017ffz}
 discussed in the appendix and current estimates for the number of flux vacua 
 \cite{Taylor:2015xtz} give:
 \begin{center}
Memory required for string geometries $\geq 10^{745}$ GB \\ \qquad \\ 
Memory required for flux vacua $\geq O(10^{272,000})$ GB.
 \end{center} 
While one of these numbers is clearly much larger, we mention the smaller one because it
arises from an exact lower bound on the number of topologically distinct geometries possible in string 
compactifications\footnote{Under the assumption that a consistent flux can be turned on in
an $O(1)$ fraction of the geometries. In fact, each geometry likely admits many consistent
fluxes, since it is essentially putting a Bousso-Polchinski \cite{Bousso:2000xa} type
problem on top of each geometric choice. This is an interesting direction for further
study.}, i.e., it is an extremely large number from an exact combinatoric calculation.
Just as physicists know that the GUT scale and Planck scale are ``large'' in the
sense discussed, we know that these datasets are large: it is highly unlikely that any
computer will ever be able to hold them in memory. 

Another interesting possibility, a priori, would be to use so-called streaming algorithms to
systematically read through the data and keep track of certain properties, without
storing it in memory. However, under the assumption that reading each data point in the
stream takes at least a Planck time, which is $O(10^{-61})$ $T_{\text{univ}}$, we obtain
 \begin{center}
Time required for streaming string geometries $\geq 10^{684}$ $T_{\text{univ}}$ \\ \qquad \\ 
Time required for streaming flux vacua $\geq O(10^{272,000})$ $T_{\text{univ}}$,
 \end{center}
 where
$T_{\text{univ}}\simeq 10^{10}$ years is the age of the universe. Clearly both memory
issues and time issues come to the fore, as of course one would have expected.

 For these reasons a
 statistical approach \cite{Douglas:2003um} to the landscape seems critical, and a 
 natural question is how to study statistics (including possible vacuum and / or anthropic
 selection) of a set of vacua that cannot be read into memory or scanned over, particularly
 given that the difficulty of the problem is exacerbated by 
 computational complexity \cite{Denef:2006ad,Cvetic:2010ky,Denef:2017cxt}. 

 One possibility
 is that some large ensembles may allow for a high degree of control, including
  potential demonstration
 of universality, due to the existence of a concrete construction algorithm from which statistical properties can be extracted; see \cite{Halverson:2017ffz}. This technique
 is clearly useful when applicable, since analytic results are almost always easier to
 understand than numeric results. However, many problems may not admit such a concrete
 construction algorithm.

 Using techniques
 from data science that have revolutionized other fields, such as supervised machine
 learning, genetic algorithms, or deep reinforcement learning, may also be a promising direction; see
 \cite{Abel:2014xta,He:2017aed,Krefl:2017yox,Ruehle:2017mzq,Carifio:2017bov} for initial studies in string theory. One might
 object that such techniques cannot be used on datasets that cannot fit into computer
 memory, but in fact this is not necessarily true, depending on the task at hand. For
 example, AlphaZero \cite{alpha-zero} is the state of the art Go program that recently
  defeated AlphaGo Lee $100$-$0$, the program that itself defeated the human world champion for
  the first time, in 2016. It did so with under $48$ hours of training for
  its neural networks, with no human knowledge input into the training (i.e.,
  training only with self-play), despite the fact that Go has $1.4\times 10^{172}$ board
  positions, far more than can fit into memory. The situation is helped by the fact that
  the ``dynamics'' of playing the game well accesses a small subset of the board positions,
  but it may also be so with the dynamics of the landscape. Furthermore, in a concrete
  example in string theory (the $10^{755}$ ensemble) it is already the case \cite{Carifio:2017bov} that machine
  learning has led to a statistical result that had not been arrived at by analytic means.
  In that case, study of the machine learned model led to a conjecture that was turned
  into a rigorous analytic theorem.

  It would be interesting to use any of these techniques to make further concrete statements about
  string remnants across across such large (in the sense discussed) datasets in the landscape.

\section{Typical Stringy Remnants}\label{remnants}
In this section we survey  typical stringy remnants  frequently encountered in semi-realistic constructions, commenting briefly on their theory, phenomenology, and implications.

\subsection{Moduli and Axions}

Scalar moduli and pseudoscalar axions  are two of the most common remnants that arise in
string compactifications. For additional discussion related to the content of this
section, we refer the reader to appendices \ref{app:axions} and \ref{app:moduli}.

Moduli
are often (but not always) associated with metric deformations of the extra dimensional
manifold that preserve a property related to supersymmetry; for example, the 
Calabi-Yau condition of a Ricci-flat K\"ahler metric on a six-manifold. Axions typically
arise from Kaluza-Klein reduction of higher form gauge fields, and the gauge symmetry of
the higher form gauge field becomes a shift symmetry of the axion that survives to all
orders in perturbation theory.

The relevance of moduli and axions  arises from the fact that they couple to many
fields in the low energy effective action, and therefore the process that cosmologically
reheats the visible sector (such as inflationary reheating) likely also transfers some
energy density into moduli or axions. Both possibilities may involve deviations from standard
cosmological paradigms that were arrived at from the bottom-up.

The typicality of moduli and axions follows in part from concrete properties of compactification manifolds,
with the implication that there are usually dozens or hundreds of moduli and~/~or axions arising
in a string compactification. 

\subsubsection{Moduli Domination}

If reheating transfers energy into the moduli sector, on long time-scales the modulus will 
begin to oscillate around its minimum, in which case it behaves like non-relativistic 
matter. Then the energy density of the modulus
$\rho_\phi$ scales as 
\begin{equation}
\rho_\phi \propto \frac{1}{a(t)^3},
\end{equation}
where $a(t)$ is the scale factor of a Friedmann-Robertson-Walker (FRW) cosmology. In the standard
cosmological paradigm reheating gives rise to a phase of radiation domination that carries
through big bang nucleosynthesis (BBN) and eventually gives rise to matter domination at the
time of matter-radiation equality (MRE).
However, a reheated modulus will often come to dominate the energy density of the universe
prior to MRE
due to its scaling relative to the scaling $\rho_{rad}\propto 1/a^4$ of radiation
in the visible sector. 
This early phase of matter domination due to moduli domination
spoils the successful predictions of BBN if the universe is matter dominated at the would-be BBN time; 
this is known as the cosmological moduli problem \cite{Coughlan:1983ci,deCarlos:1993wie,Banks:1993en}.
This is avoided in string-motivated scenarios where the modulus decays via a dimension-five Planck-suppressed
operator if $m_\phi \gtrsim 50$ TeV.

There are relatively few constraints on such a phase of matter domination prior to nucleosynthesis, and
if it arises due to modulus domination new cosmological paradigms arise, such as the production of
dark matter via modulus decay (e.g., \cite{Moroi:1999zb,Acharya:2008bk,Acharya:2009zt,Fan:2013faa,Cohen:2013ama}) rather than thermal freeze-out. Observables of inflationary models can be modified depending
on whether they have a standard or moduli dominated cosmology prior to BBN \cite{Easther:2013nga}.

See the review \cite{Kane:2015jia} and appendix \ref{app:moduli} for further discussion.
See also \cite{Antusch:2017flz} for recent work on localized, long-lived, non-linear excitations of moduli known as oscillons.

\subsubsection{Axion Inflation, Dark Matter, and Dark Radiation}

Broadly, axions or axion-like particles (ALPS) are to be expected in string compactifications, forming
a so-called axiverse \cite{Arvanitaki:2009fg}. 
Their basic phenomenology is determined by their couplings, decay constants, and masses.
Distributions of axion decay constants tend to be centered an order of magnitude or two
below $M_{p}$, while their masses may take a variety of values; see, e.g.,~\cite{Long:2014fba,Stott:2017hvl}. Accordingly, they may play a number of qualitatively different roles, and we categorize
the possibilities according to three such roles.

Models of inflation where the inflaton is an axion have been extensively studied. Axions are one of the
few particle types that may successfully give rise to large-field inflation in string theory, due to
the extremely sensitive dependence on large-field inflation on Planck-suppressed operators. These operators
are forbidden in perturbation theory by the axion shift symmetry, which plays a critical role in
the viability of the models. Two basic categories of models are referred to as axion inflation 
(see, e.g.,~\cite{Svrcek:2006yi,Grimm:2007hs,Long:2014dta}), where
the inflationary potential is a standard cosine in the axion field, and axion monodromy 
\cite{Silverstein:2008sg,McAllister:2008hb,Marchesano:2014mla}, which
includes extra terms such as a linear term in the axion potential.

When axions oscillate around their minimum they behave as non-relativistic matter, in which
case they may be dark matter candidates. 
They may behave as standard QCD axion dark matter, in which case the  axion couples strongly to the gluon field strengths and is produced by
misalignment. However, the upper bound on the QCD axion decay constant $f<10^{12}$ GeV\footnote{This constraint can be weakened in some cases if the initial value of the axion field is fortuitously close to the minimum.  See, e.g.,
\cite{Wilczek:2004cr}.}
is several orders of magnitude smaller than typical decay constant scales in string theory.  
 Alternatively, many of the axions in string theory do not behave as the
QCD axion, and they may comprise some or all of the dark matter. See \cite{Marsh:2015xka}
or appendix \ref{app:axions} for a discussion of the possibility that string axions comprise
so-called fuzzy dark matter, which may account for some astrophysical
anomalies such as the core vs. cusp problem.

Axions may also behave as dark radiation, in which case it is constrained by bounds on $\Delta N_{\text{eff}}$. See \cite{Higaki:2012ar,Cicoli:2012aq,Cicoli:2015bpq} for studies of axionic dark radiation
in LARGE volume compactifications. Axion-photon scattering during BBN can affect light element
abundances \cite{Conlon:2013isa}, and axion dark radiation can constrain neutralino dark matter when it is produced by modulus decay, due to correlations between the physics
of axions and moduli \cite{Allahverdi:2014ppa}. Considerations from the string axiverse imply
that a cosmic axion background exists today; see \cite{Conlon:2013isa}. \jh{However,
axions arising from the decay of moduli
that solve the cosmological moduli problem may still be significantly constrained \cite{Higaki:2013lra}.}

We refer the reader to~\cite{Marsh:2015xka} for a review of axion cosmology, to~\cite{Baumann:2014nda} for
a review on inflation in string theory that includes axions, and also to references therein.

\subsection{Extended Gauge Sectors}

\subsubsection{TeV-Scale {\zpr}s and other Extended Gauge Symmetries}\label{zprime}
Many types of new physics lead to the existence of new gauge bosons at the TeV scale or higher~\cite{Langacker:2008yv}.
{\zpr}s are especially common because it is often harder to break {\uprm}s than non-abelian symmetries, i.e.,
they tend to ``slip through'' when a larger non-abelian symmetry is broken. They frequently occur not only as string remnants, but also in alternative models of electroweak breaking such as dynamical symmetry breaking or composite Higgs models,  and  in the breaking of grand unified theories larger than \stf.

In the string context (e.g.,~\cite{Faraggi:1989ka,Faraggi:1990ita,*Cleaver:1997rk,*Faraggi:2015iaa,Cvetic:1997wu,*Cvetic:1996mf,Cleaver:1998gc,
Cvetic:2001tj,*Cvetic:2001nr,Blumenhagen:2005mu,Anastasopoulos:2006da,Cvetic:2011iq}) additional non-anomalous {\uprm}s often emerge as unbroken subgroups of non-abelian groups such as 
$SO(10)$~\cite{Robinett:1981yz}, $E_6$~\cite{Hewett:1988xc}, or the $SU(4)\x SU(2)_{L} \x SU(2)_{R}$  Pati-Salam group~\cite{Pati:1974yy}  (any of which can be broken in the extra dimensions).
The \uprm\ charges are then typically linear combinations of  $T_{3R}$, $B-L$, or the $E_6$ charges. Additional  {\uprm}s may instead be ``random'', i.e., not resembling familiar grand unification or other extended gauge symmetries.
In the latter case the associated {\zpr}s may couple to particles in a hidden or quasi (partially)-hidden sector,
to some or all of the standard model particles, or both, and they can sometimes
 mediate transitions between the sectors by direct couplings or by kinetic mixing. The couplings are often family-nonuniversal, leading to flavor-changing effects.
In many cases  anomalies are canceled by new quasi-chiral fermions (i.e., fermions that are vector-like under the standard model gauge group).

{\uprm}s may also be ``anomalous'' (in the traditional field theory sense~\cite{Adler:1969gk,*Bell:1969ts}), although the apparent anomalies are actually canceled by
Chern-Simons terms to yield consistent theories~\cite{Ibanez:1428137,Dine:1987xk,*Atick:1987gy,Kiritsis:2004kca,*Coriano:2005own,*Anastasopoulos:2006cz}.\footnote{Such four-dimensional field theories are sometimes referred to as augmented field theories.} For example,  heterotic constructions often include one anomalous \uprm. They are especially prevalent in Type II string theories, which involve $U(N)$ gauge groups  rather than $SU(N)$
[$Sp(2N)$ and $SO(2N)$ are also possible], with the trace $U(1)$  usually anomalous. 
The  gauge boson associated with an anomalous  $U(1)$  generally acquires a string-scale mass by the  St\"uckelberg mechanism~\cite{Stueckelberg:1900zz} 
and is therefore irrelevant except in the special case of a low string scale. However, the \uprm\ symmetry survives as a global symmetry on the perturbative couplings of the low energy theory, possibly broken by non-perturbative string instanton effects~\cite{Blumenhagen:2006xt,Ibanez:2006da,Blumenhagen:2009qh,*Cvetic:2011vz}.
Moreover, one or more linear combinations of the anomalous {\uprm}s  may be non-anomalous due to (accidental) cancellations, with the associated {\zpr}s emerging as ordinary gauge bosons that can acquire mass by the Higgs or other  mechanisms. The standard model hypercharge often emerges in this way, and other cancellations can lead to remnant  {\zpr}s. The latter are often family-nonuniversal.

\subsubsection*{Experimental signatures of a \zpr}

A \zpr\ can be searched for at the LHC~\cite{Langacker:2008yv,Olive:2016xmw}, e.g.,  through Drell-Yan $pp\ra \zpr\ra \mu^+\mu^-,e^+e^-,$  $q\bar{q}$, $W^+W^-$, or $ZH$.
Limits from early LHC running exclude masses below 3-4 TeV for electroweak-strength couplings to ordinary fermions, and should eventually reach 5-6 TeV, with the specific value dependent on the \zpr\ gauge coupling, charges, and the effects of decays into exotics and superpartners. Diagnostic studies of the charges, $Z-\zpr$ mixing, etc.,  could in principle be carried out by measurements of branching ratios,  asymmetries, polarizations, rapidity distributions, associated production,
 and rare decays, but in practice this would be difficult above 3-4 TeV. Masses up to  $\mathcal{O}(30 \text{ TeV})$
 could be probed at a  future  100 TeV $pp$ collider~\cite{Arkani-Hamed:2015vfh}. {\zpr}s could also be searched for (and their couplings probed)  at a future $e^+e^-$ collider, where interference effects with $s$-channel $Z$ and $\gamma$ exchange should be sensitive to masses well above $\sqrt{s}$~\cite{Han:2013mra,Moortgat-Picka:2015yla}. Existing $Z$-pole and weak neutral current results typically limit any $Z-\zpr$ mixing angles  
 $|\theta_{Z-Z'}|$ to less than a few $\x  10^{-3}$. 
 
 Lighter {\zpr}s, e.g., in the 10-1000 GeV range, with couplings much smaller than electroweak or with special couplings (e.g., leptophobic) are also possible~\cite{Jaeckel:2010ni,*Goodsell:2009xc,Essig:2013lka,Alexander:2016aln}. These could be searched for in high precision weak neutral current experiments,  rare decays, or high-luminosity LHC running. Very light (e.g., $< 10$ GeV or even much smaller) {\zpr}s are sometimes invoked in connection with hidden sector dark matter models, often serving as a portal between the dark and ordinary sectors. Their couplings to the ordinary particles would have to be extremely weak, e.g., generated only by kinetic mixing. Very light {\zpr}s can also be searched for in beam dump experiments and in astrophysics/cosmology, and could be relevant to the muon magnetic moment anomaly. 
 
\subsubsection*{Other implications  of a TeV-scale \zpr}
The existence of a new \zpr\ (whether or not associated with string theory) would have many other implications. Here we mainly consider 
a \uprm\ with electroweak couplings broken at the (multi-) TeV scale. For more detail see~\cite{Langacker:2008yv}.

\begin{itemize}
\item The \uprm\ charges could give clues as to the embedding into the underlying theory.

\item A \uprm\ could give a natural solution~\cite{Suematsu:1994qm,Cvetic:1997ky} to the $\mu$ problem~\cite{Kim:1983dt} of the MSSM.\footnote{Other solutions include the Giudice-Masiero mechanism~\cite{Giudice:1988yz} and string instantons~\cite{Blumenhagen:2006xt,Ibanez:2006da}.} That is, if the two Higgs doublets $H_{u,d}$ of the MSSM do not form a vector pair then an elementary $\mu H_u H_d$ term in the superpotential
would be forbidden, but an effective $\mu_{eff} = \lambda_S \langle S \rangle$ would be generated if $ \lambda_S  S H_u H_d$ is allowed, where $S$  is a standard model singlet\footnote{We use the same symbol for both the chiral superfield and its scalar component.} with nonzero charge.  $\langle S \rangle$ also generates or contributes to the \zpr\ mass and may generate masses for exotics.
This can be viewed as a stringy version of the NMSSM~\cite{Maniatis:2009re,Ellwanger:2009dp}, except there is no discrete $Z_3$ symmetry or attendant cosmological domain wall problems. Also, the MSSM bound of around 135 GeV on the lightest Higgs scalar mass is increased to $\sim$150 GeV by new $F$-term (also present in the NMSSM) and $D$-term contributions to the scalar potential~\cite{Barger:2006dh}.

\item More generally, but assuming that the $\mu$ problem is somehow solved,  the electroweak and \uprm\ breaking scales are typically both set by the  soft supersymmetry breaking and $\mu$ scales. Thus, the \zpr\ and related exotic masses are usually not too much larger than the superpartner masses, or to the electroweak scale, up to an order of magnitude or so. This conclusion can be evaded, however, if the \uprm\ is broken along a flat or almost flat direction~\cite{Cleaver:1997nj,*Erler:2002pr} or by fine-tuning.

\item
\uprm\ models typically involve one or more standard model singlet fields $S$ whose expectation values break the symmetry,
implying at least one additional Higgs scalar. 
The extra scalars are often at the TeV scale, but in principle could be lighter. 
They could mix with the standard model or MSSM Higgs bosons, modifying couplings and their experimentally-allowed ranges~\cite{Barger:2006dh}.

\item Supersymmetric \uprm\ models have an extended neutralino sector~\cite{Barger:2006kt}, including at least the \zpr\ gaugino ($\tilde Z'$ )
and the singlino ($\tilde S$), modifying couplings or allowing new possible scenarios for collider physics (e.g., extended cascades ending in the production of a light invisible singlino  and a $Z$ or $H$), cold dark matter, etc.

\item Non-anomalous {\uprm}s often involve quasi-chiral exotic fermions that cancel anomalies. Familiar examples are
\beq D_L + D_R, \qquad \vect{E^0 \\ E^-}_L+ \vect{E^0 \\ E^-}_R,
\eeql{E6exotics}
where $D_{L,R}$ are an \st-singlet pair of quarks with electric charge $-1/3$, and $(E^0 E^-)_{L,R}^T$ are an \st-doublet
pair of leptons or Higgsinos.
These occur in the 27 of $E_6$ along with a standard model family, a right-handed neutrino $\nu_R$, and a singlet $s$ whose scalar partner \veva{S}\ can break the \uprm, and are also frequently encountered as string remnants.  Their properties are further discussed below.

\item The decays of a  heavy \zpr\ could be an efficient means of producing superpartners, exotics, heavy right-handed neutrinos, etc.
(e.g.,~\cite{Baumgart:2006pa}).

\item Stringy \zpr\ couplings are often family-nonuniversal, e.g., because the  standard model families have different
string constructions,  because of Kaluza-Klein excitations coupling to  fermions with different spatial distributions in large or warped extra dimensions~\cite{Delgado:1999sv}. 
Nonuniversal couplings usually lead to flavor-changing vertices when fermion mixing is turned on~\cite{Langacker:2000ju}. The limits  from rare decays and $K_L-K_S$ mixing are sufficiently strong that such nonuniversality is probably restricted to the third family. It could possibly be associated with the observed hints of nonuniversality in $\Gamma(B\ra K\mu^+\mu^-)/\Gamma(B\ra Ke^+e^-)$ (see, e.g.,~\cite{Ellis:2017nrp,Altmannshofer:2017yso,Crivellin:2017ecl}),
could lead to new tree-level effects
that could compete with standard model loops (e.g.,  $B_d$ penguins or in $B_s-\bar B_s$ mixing)~\cite{Barger:2009qs,*Everett:2009cn}, or to effects in $\tau$ or $t$ decays.

\item{Additional anomalous or non-anomalous \uprm\,s can sometimes distinguish between lepton and down-Higgs doublets, which have the same gauge quantum numbers in the  ($R_P$-violating) MSSM.} They can also sometimes distinguish SM-singlet Higgs fields from sterile neutrinos.

\item A \zpr\ could serve as a portal to a (quasi-)hidden sector, either by direct coupling or kinetic mixing~\cite{Holdom:1985ag}. Similarly, a $\zpr-\tilde \zpr$ mass difference generated in a supersymmetry breaking sector could mediate supersymmetry breaking to the standard model sector~\cite{Langacker:2007ac,*Langacker:2008ip,*deBlas:2009vx,Feng:2014eja,*Feng:2014cla}.

\item An additional \uprm\ can affect the possibilities for generating small neutrino masses~\cite{Kang:2004ix}. 
Depending on the couplings, conventional models such as the Type I seesaw may be allowed or forbidden,
while novel mechanisms such as small Dirac masses by higher-dimensional operators or non-holomorphic soft terms may be allowed. The various possibilities are described in a more general stringy context below.

\item There could also be significant implications for cosmology~\cite{Langacker:2008yv}. For example,
trilinear $S H^\dagger H$ or $S H_u H_d$ Lagrangian terms  could allow a strong first-order electroweak phase transition  (as in the NMSSM), necessary for electroweak baryogenesis. Cold dark matter and big bang nucleosynthesis could be modified
in a number of ways by the extended neutralino/Higgs/\zpr\ sectors.

\end{itemize}

\subsubsection*{Non-abelian gauge symmetries at the TeV scale}
Additional non-abelian gauge symmetries are also possible, both in string theories and in
other types of standard model extension. These can easily act in hidden or quasi-hidden sectors, 
and could be either weakly or strongly coupled. One well-motivated stringy extension in the standard model
sector involves an $SU(2)$ that couples to right-chiral fermions, as in \stto\ models (see, e.g.,~\cite{Mohapatra:737303,Maiezza:2010ic,Cao:2012ng}). (There need not be any 
discrete left-right symmetry.) The phenomenological implications depend on whether the $\nu_R$ are light
(as in Dirac or pseudo-Dirac neutrino masses) or heavy (as in a seesaw model), and also on the
right-handed quark mixing matrix. These include precision measurements of $\beta$ and $\mu$ decay,
weak universality, the $K_L-K_S$ mass difference, $M_W$, and possible heavy $W'$ and $\nu_R$ production and decay.

\subsubsection{Extended Non-abelian Gauge Sectors and Hidden Sector Dark Matter}
\label{sec:extendednonabeliananddm}
Many constructions involve one or more (quasi)-hidden sectors of gauge-charged particles that
are coupled to the standard model sector via gravity and some combination of moduli, axion, and gauge-charged mediator couplings (e.g.,~\cite{Cvetic:2012kj,Heckman:2015kqk,Acharya:2016fge} and  Appendix~\ref{sec:dark gauge Ftheory}). 
We are restricting attention here to a weakly coupled standard model sector, but such hidden or quasi-hidden sectors may well involve strong coupling. 
In such a vacuum, the standard model may represent
a small fraction of the gauge-charged matter that could be excited. There are two
basic possibilities:
\begin{itemize}
\item[a)] Extended gauge sectors that couple to the standard model only via gravity, axions, and / or moduli. Such interactions with the standard model are very weak due to the fact
that the Planck scale or the string scale appears as the effective cutoff in the
associated effective operators.
\item[b)] Extended gauge sectors that interact with the standard model via a gauge-charged mediator,
	or a sequence of them if a gauge factor is separated from the standard model by multiple
	gauge factors. The strength of the interactions with the standard model depend on
	a number of factors, including the number of mediators and their masses.
\end{itemize} 
These ideas are well-exemplified in a quiver diagram, such as
\begin{center}
\includegraphics{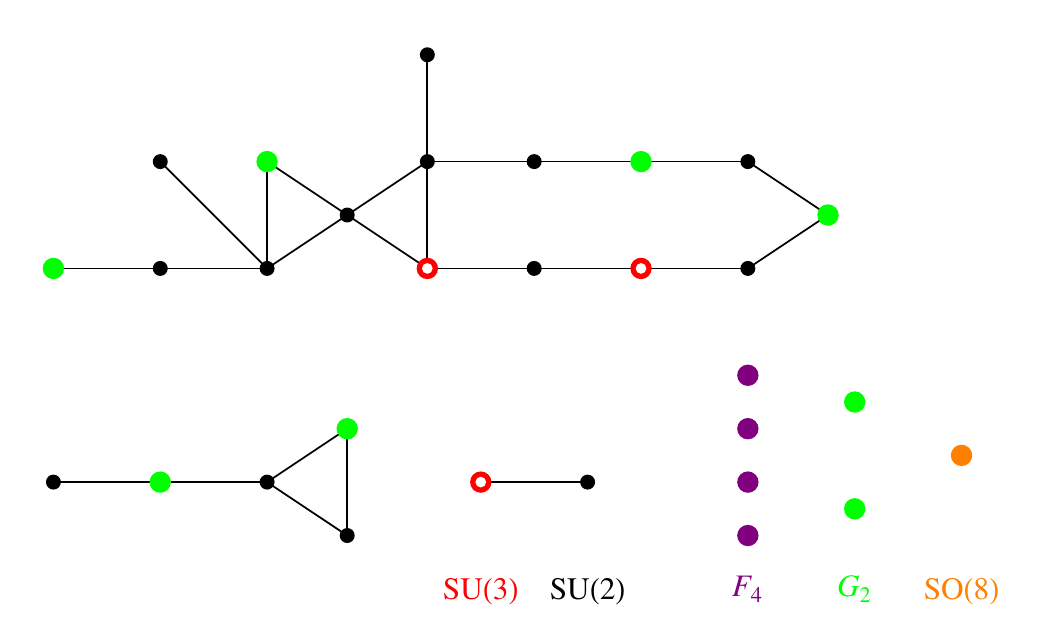},
\end{center}
which is also discussed in appendix \ref{sec:dark gauge Ftheory}. Nodes are gauge
factors that are labeled according to their group, and edges represent jointly
charged matter. 
If the compactification realizes the standard model on the
$SU(3)\times SU(2)$ factor that is disconnected from the rest of the gauge nodes,
the other gauge sectors interact with the standard model only via gravity, axions, and moduli, and are well-hidden. On the other hand, if the standard model
is realized on one of the $SU(3)\times SU(2)$ pairs in the large connected component,
then adjacent nodes may interact with it via a single mediator, while faraway nodes
are more hidden according to the number of intermediate mediators.
From this we see that some extended non-abelian gauge sectors are much more hidden than
others. In principle one can therefore distinguish between truly 
hidden sectors and quasi-hidden sectors. 

\vspace{.5cm}

There have been many phenomenological studies of hidden and quasi-hidden sectors. The most familiar
are supersymmetry breaking sectors, with supersymmetry breaking mediated to the standard model sector by gravity, anomaly, gauge, \zpr, or other mechanisms~\cite{Chung:2003fi}. 

More recently, there has been considerable interest in possible hidden  dark matter  sectors~\cite{Alexander:2016aln}, often connected by Higgs, \zpr,  neutrino, or axion portals, but
sometimes with very weak interactions with the standard model. 
These and other possibilities may arise from remnants of string theory; see, e.g., \cite{Dienes:2011ja,Acharya:2017kfi} for discussions. For example, from
extended non-abelian gauge sectors that confine, dark hadrons and / or glueballs may
arise that comprise the dark matter.
Dark
glueball dark matter \cite{Carlson:1992fn,Faraggi:2000pv,Soni:2016gzf,Forestell:2016qhc,Halverson:2016nfq,Dienes:2016vei,Soni:2016yes,daRocha:2017cxu,Acharya:2017szw,Soni:2017nlm,Forestell:2017wov} from strongly-coupled Yang-Mills sectors are possible but often problematic~\cite{Halverson:2016nfq} if the sector is hidden, rather than
quasi-hidden. \jh{Dark glueballs may also arise in warped throats \cite{Chen:2006ni,*Harling:2008px},
which occur frequently (see, e.g., \cite{Hebecker:2006bn}) in the landscape.} Another possibility is that that dark matter is dynamical \cite{Dienes:2011ja,Dienes:2011sa},
existing in multiple components with a variety of
lifetimes and abundances that are balanced against one another.

Hidden or quasi-hidden sectors (e.g.,~\cite{Cleaver:1998gc,Cvetic:2002qa}) could also be random and not connected to either supersymmetry breaking or dark matter. These would not necessarily have any significant observational consequences, but in some opportune cases there could be ``hidden valley'' signatures such as lepton jets, displaced vertices, and missing energy~\cite{Strassler:2006im,*Han:2007ae}.

\subsection{Particles Beyond the Standard Model}

\subsubsection{Extended Higgs/Higgsino sectors}
Extended  Higgs/Higgsino sectors are quite common, with  implications for collider physics, $CP$ violation, flavor physics, and cosmology.

 Standard model singlets $S$ have already been mentioned
in connection with additional {\zpr}s, but occur much more generally~\cite{Anastasopoulos:2006da,Cvetic:2010dz,Cvetic:2011iq}. Often they can acquire vacuum expectation values
and possibly can mix with Higgs doublets. These can break \uprm\ and other symmetries, give masses to quasi-chiral exotics
(including the $H_{u,d}$ pair, solving the $\mu$ problem and allowing a strong first-order electroweak transition),
and can occur in higher-dimensional operators (e.g., to generate suppressed Yukawa couplings). Other
singlets can have couplings and quantum numbers that allow them to  play the role of right-handed or other sterile neutrinos, or to be  dark matter candidates (e.g.,~\cite{Barger:2008jx}). Yet others can be unrelated to either neutrinos or symmetry breaking.

Also very common are extra Higgs doublets, especially vector-like  pairs in addition to the single $H_{u,d}$ pair of the MSSM.
These and the associated Higgsinos obviously lead to a much richer spectrum and collider phenomenology.
They  may also be needed for or expand the possibilities for describing quark and lepton masses and mixings.
For example, stringy symmetries that differentiate between the fermion families would restrict the perturbative couplings of a single Higgs doublet, so that different elements of the fermion mass matrix could be associated with different doublets. This would allow some of all of the fermion mass hierarchies to be due to hierarchies in the expectation values of the multiple Higgs fields rather than in the Yukawa couplings. However, such models also open up the possibility of flavor changing neutral  current effects mediated by neutral Higgs fields~\cite{Glashow:1976nt}. Another difficulty is that extra Higgs doublets, e.g., an extra $H_{u,d}$ pair added to the MSSM, would significantly modify gauge coupling unification unless the doublets are accompanied by other
new particles, e.g., to form complete  5-plets of \stf.

Higgs triplets, especially those with hypercharge 0 [i.e., $(\phi^+, \phi^0, \phi^-)^T$] can also occur.

See the Higgs Boson review in~\cite{Olive:2016xmw} for a detailed discussion of extended Higgs sectors in a general context.

\subsubsection{Quasi-Chiral Exotics}

One frequently encounters exotics that are non-chiral with respect to the standard model gauge group
but chiral under extended gauge, (perturbative) global, or discrete symmetries~\cite{Anastasopoulos:2006da,Cvetic:2011iq,Halverson:2013ska}.
Such exotics are often required to cancel anomalies for non-anomalous extended gauge symmetries,
or due to string consistency conditions that do not have a simple analog in
quantum field theory (see appendix \ref{app:consistency} for a discussion of the latter).
They may couple directly to the standard model sector through gauge or Yukawa interactions, but could also be restricted to a hidden sector. If they are charged under the standard model gauge group they generally modify gauge unification unless they occur in full \stf\ multiplets (similar to the comment on additional Higgs doublets).

 Common examples of quasi-chiral exotics  include the isosinglet down-type quark and isodoublet lepton/Higgsino vector-like pairs\footnote{Whether these are best interpreted as exotic leptons, as additional Higgs pairs, as both, or as neither, depends on the allowed couplings and symmetry breaking. Similar statements apply to neutral singlet fields, which may play the role of Higgs particles ($S$), sterile neutrinos ($N^c$), both, or neither.} in \refl{E6exotics}. These can also be written as left-chiral
 supermultiplets,
\beq D + D^c, \qquad \vect{E^0 \\ E^-}+ \vect{E^+ \\ E^{0c}}.
\eeql{E6exotics2}
Other commonly encountered pairs
are  isosinglet up-type quarks $U+U^c$, isodoublet quarks $(U D)^T+(D^c U^c)^T$, non-abelian singlets like $E^-+E^+$ with 
$Y=Q=\mp 1$,  and shifted-charge  lepton/Higgsino doublets  with charges $(\pm1,\pm2)$.
Yet other types of pairs, such as color sextets, isosinglet charge $-4/3$ quarks, or  isodoublet quarks with shifted charges $(-1/3,-4/3)$ are also in principle possible,
but did not occur in  the quiver survey in~\cite{Cvetic:2011iq}. The possibility of fractionally-charged color singlets will be discussed separately below.

The scalar components of exotics can acquire masses by soft supersymmetry breaking. However, mass generation for the fermions is sometimes problematic. If the chiral symmetry is broken perturbatively by the expectation values of a standard model singlet $S$ then the fermions (and the scalars if the breaking scale is larger than the supersymmetry scale)
will typically 
acquire masses related to the symmetry breaking scale. For example, exotic pairs $X+X^c$ would become massive
through couplings like $S X X^c$ or by higher-dimensional operators involving $S$ if they are perturbatively allowed.
Masses could also be generated by nonperturbative breaking of the chiral symmetries, e.g., by string instantons
\cite{Blumenhagen:2006xt,Ibanez:2006da}.
 In other vacua that are not consistent with our own the fermions may remain massless.

Vector pairs that are non-chiral  under all symmetries are a common special case. At generic
points in moduli space these would often acquire string-scale masses and are therefore irrelevant for our purposes unless the string scale is low. However, they may become light at subloci in moduli space when the combination
of field expectation values that determines the mass of the vector pair goes to zero.

\sto\ singlets such as $S$   and \st-triplets with $Y=Q=0$ can also occur. These are non-chiral under \sto\ and can occur either in pairs or singly. The fermions could in principle acquire either Dirac or Majorana masses.

Of course, string vacua can also involve additional matter that is standard model chiral but anomaly free. The most obvious possibility is a chiral sequential or mirror fourth family. Another is the shifted family
\beq
(\textbf{3},\textbf{2})_{-\frac{5}{6}},\ \
		(\ov{\textbf{3}},\textbf{1})_{\frac{1}{3}},\ \
		(\ov{\textbf{3}},\textbf{1})_{\frac{4}{3}},\ \
		(\textbf{1},\textbf{2})_{-\frac{3}{2}},\ \
		(\textbf{1},\textbf{3})_{1},
		\eeql{shiftfourth}
where the numbers in parentheses are the \sth\ and \st\ representations, and the subscript is $Y$.
However,  exotics that are chiral under  \sto\ would have to acquire masses by electroweak breaking. \jh{Experimental limits are sufficiently strong that they would require very large Higgs-Yukawa couplings, leading to Landau poles at low energy,
violation of tree-level unitarity, and probable conflict with the observed Higgs couplings and electroweak precision tests.}

The phenomenology of vector-like exotics has been extensively discussed.\footnote{A partial list includes~\cite{Hewett:1988xc,Langacker:1988ur,Choudhury:2001hs,Kang:2007ib,delAguila:2008pw,Endo:2012cc,Martin:2012dg,Botella:2012ju,ArkaniHamed:2012kq,Kearney:2012zi,Batell:2012ca,Joglekar:2013zya,Garberson:2013jz,Aguilar-Saavedra:2013qpa,Fairbairn:2013xaa,Altmannshofer:2013zba,Halverson:2014nwa,Ellis:2014dza,Ishiwata:2015cga,Alloul:2013raa,Cvetic:2015vit,Lalak:2015xea,Kumar:2015tna,Bobeth:2016llm,Joglekar:2016yap,Aguilar-Saavedra:2017giu,Poh:2017tfo}.} Here we mainly follow the supersymmetric treatment in~\cite{Kang:2007ib}.
Colored exotic quarks and squarks could be  efficiently pair produced at a hadron collider by QCD processes if their masses are not too large, while color singlets would have much lower rates. Single production by mixing  or leptoquark or  diquark couplings would be kinematically favored but with smaller couplings. The relative masses of the quark and two squarks are dependent on the $D$-terms and soft breaking, with the heavier ones cascading downwards by processes like $\tilde D_1 \ra D \tilde G$ or $D\ra \tilde D_1 \tilde \chi^0$, where $\tilde G$ and $ \tilde \chi^0$ represent  gluino and neutralino, respectively. 

The exotics might be able to  decay into standard model particles by mixing. For example, an isosinglet down-type quark
might decay into $D\ra u_iW^-,$ $D \ra d_i Z,$ or $D \ra d_i H^0$ by $D-d_i$ mixing. However, the extra symmetries often forbid or strongly suppress this. Another possibility is decays via allowed  leptoquark or diquark  operators,
such as $D u^c e^c$, or $D^c u^c d^c$, respectively.\footnote{Unlike the possible leptoquark or diquark operators in the MSSM, these exotic operators still allow for a conserved $R$-parity and stable LSP.}
Of course, proton stability requires that they should not both be present simultaneously at any significant level. The lightest exotic could also be stable at the renormalizable level, decaying only by higher-dimensional operators or nonperturbative effects. Depending on the details, 
exotics could decay promptly; with displaced vertices; pass out of the detector; or hadronize and stop, decaying later (a form of $R$-hadron~\cite{Kraan:2004tz,*Kraan:2005ji}). They could even be approximately stable on cosmological time scales, although that possibility is constrained by big bang nucleosynthesis.

 Other possible implications of quasi-chiral exotics include flavor, precision, and Higgs physics; flavor-changing and nonuniversal effects;  gauge mediation; and cosmology.

\subsubsection{Absence of Large Representations}
Perturbative string constructions typically do not allow large or unusual 
 representations for the elementary
matter multiplets in the low energy four-dimensional  gauge theory.
For example, simple heterotic constructions
only allow singlets,  fundamentals (or bifundamentals under two groups), and their conjugates.
In some cases this constraint can be evaded if the  group is realized at a higher  Ka\v{c}-Moody level~\cite{Lewellen:1989qe,Dienes:1996yh,*Dienes:1996wx}. 
For example, the diagonal subgroup   $G \in G \x G$ corresponds to a level 2 embedding, and can yield
adjoints, symmetric and antisymmetric representations, and their conjugates. However, higher embeddings are
increasingly complicated and there are restrictions on their size.

Similarly, 
 Type IIA intersecting brane constructions are restricted to singlets, fundamentals, bifundamentals, adjoints, two-index
  symmetric and antisymmetric representations, and their conjugates.
 Large representations, trifundamentals, etc., may be possible in F-theory and M-theory, but
 relatively little investigation has been done on the subject. Large gauge groups are to be
 expected in F-theory due to the prevalence of non-Higgsable clusters \cite{Taylor:2015ppa,Halverson:2017ffz,Taylor:2017yqr}, however. Since they correlate strongly with being
 at strong string coupling \cite{Halverson:2016vwx}, that is, in moving away from
 weakly coupled type IIB, it may be that such effects are also generic in M-theory,
 that is, in moving away from weakly coupled type IIA.

\subsubsection{Fractionally Charged Color Singlets}

Fractional electric charges for color singlet particles or bound states 
are  common in string constructions, e.g.,~\cite{Wen:1985qj,Faraggi:1990af,*Chang:1996vw,*Coriano:2001mg,Lykken:1996kc,Dienes:1995sq,Chaudhuri:1995ve,Cleaver:1998gc,Cvetic:2001nr,Cvetic:2002qa,Cvetic:2011iq}, and constitute a problem rather than a signature. These may, for example, be associated with \st\  singlets with hypercharge $Y=\pm\oh$, nonstandard $Y$, or additional contributions to the electric charge generator $Q$ for exotics.\footnote{Small, e.g., millicharge,
particles can also be generated by kinetic mixing, such as between the photon and a hidden-sector dark boson~\cite{Holdom:1985ag}. However, this is not necessarily a stringy effect.}  
For example, the Schellekens theorem~\cite{Schellekens:1989qb} states that otherwise-realistic weakly coupled heterotic constructions require either fractional charges, an unbroken \stf, or higher-level Ka\v{c}-Moody embeddings for \st\ or \sth.

The lightest fractionally-charged particle would be stable for an unbroken $Q$, and (except for very small charges) are  excluded for almost all masses by direct searches~\cite{Perl:2009zz} and cosmological constraints such as overclosing the universe~\cite{Langacker:2011db}. However, these constraints can be evaded if  the particles are too heavy to produce following inflation or are confined by additional non-abelian gauge interactions.

\subsection{Couplings and Hierarchies}

\subsubsection{Leptoquark, Diquark, Dilepton, and $R_P$-Violating Couplings}

String constructions can allow perturbative leptoquark, diquark, or dilepton couplings between new quasi-chiral exotics and 
 ordinary particles.
Of course, the simultaneous presence of both leptoquark and diquark couplings would lead to rapid proton decay, but one or the other by itself would be possible. There are numerous implications for collider physics and for flavor changing or nonuniversal processes~\cite{Barbier:2004ez,Dorsner:2016wpm}, such
as the hints of anomalies in $B$ decays (e.g., $\Gamma(B\ra K\mu^+\mu^-)/\Gamma(B\ra Ke^+e^-)$~\cite{Hiller:2017bzc}, or 
$\Gamma(B\ra D^{(\ast)} \tau \nu)/\Gamma(B\ra D^{(\ast)} \ell \nu)$~\cite{Altmannshofer:2017poe}).
 Other collider implications  include decay mechanisms for quasi-chiral exotics
that are sometimes forbidden to significantly mix with ordinary particles~\cite{Kang:2007ib}. 

The extended (beyond-MSSM) symmetries can either allow or forbid $R_P$-violating couplings
 amongst standard model particles. For example,  $H_d$ and the standard model lepton doublets $L$ have the same standard model gauge assignments. If, however, they carry different extended quantum numbers then the couplings
 $H_u L$,  $LLE^c$, and  $LQD^c$
 will be forbidden, at least at the perturbative level. The diquark coupling $U^cD^cD^c$ may be similarly forbidden.
 Conversely, in some vacua some of these couplings are allowed,
 implying   $R_P$ and    $L$ and/or $B$ violation.

\subsubsection{Family Nonuniversality}
Family nonuniversality, typically because of different stringy origins of the fermion families, has already been mentioned
in connection with \uprm\ couplings  and with selection rules for Higgs couplings. Such effects can lead to flavor changing neutral currents or 
could be relevant to understanding fermion mass hierarchies.
Mixing between ordinary and exotic fermions with different \sto\ assignments will in general also lead to flavor changing effects involving the ordinary $Z$~\cite{Langacker:1988ur}.

\subsubsection{Mechanisms for Yukawa Hierarchies}
String theories admit  a number of possible origins of fermion masses and mixings (as well as other couplings, such as those leading to quasi-chiral exotic masses), and it is possible or even likely that the details of the fermion spectrum are to some extent accidental features of the location in the landscape, and not susceptible to any simple explanation.

A perhaps simpler issue is the origin of the observed large hierarchies of masses and mixings. Here, there are again various possibilities.
Let us consider two very different examples.

 One possibility is that the Yukawa and related couplings actually result from  stringy higher-dimensional  operators.\footnote{By stringy operator we mean that the inverse masses are due to the underlying string dynamics rather than  the exchange of a massive particle in the four-dimensional field theory.} 
For example, the various $u$-type quark Yukawas couplings 
\beq W_U = \Gamma^u Q H_u U^c\eeql{uyukawa}
could actually result from couplings like
\beq  \frac{1}{M_s^n} Q H_u U^c S_1 S_2 \cdots S_n,\eeql{uhdo}
where $S_i, i=1\cdots n$, are standard model singlets  (some of which could be the same) and $M_s$ is the string scale.
Effective couplings  of various magnitudes could arise if the $S_i$ obtain expectations values smaller than $M_s$.
For example, the $t$, $c$, and $u$ Yukawas could be associated with operators with $n=0,$ 2, and 5, respectively,
and $\veva{S_i}/M_s\sim 0.1$. This sort of thing often occurs in heterotic constructions (e.g.,~\cite{Cleaver:1998gc}), with the selection rules on the allowed operators controlled by additional gauge  symmetries. The magnitudes of the \veva{S_i}\   are obtained by minimizing the scalar potential in  the presence of a non-zero Fayet-Iliopoulos term for an anomalous \uprm,
a process known as vacuum restabilization~\cite{Dine:1987xk,*Atick:1987gy}.\footnote{The restabilization also reduces the gauge symmetry and generates string-scale masses for some of the exotics.}
 This mechanism is basically a stringy version of the Froggatt-Nielsen mechanism~\cite{Froggatt:1978nt}.

In Type IIA intersecting brane constructions, on the other hand, chiral fermions are trapped at the intersections of D6 branes. 
Consider, for simplicity, the case in which the extra dimensions form three 2-tori. Three-point couplings  result when three stacks of branes form  triangles  of  area $A_I$ in the $I^{th}$ torus, with the Yukawa couplings  proportional to $\prod_I \exp{(-A_I/2\pi\alpha')}$, where $\alpha'$ is the string tension~\cite{Aldazabal:2000cn,Cvetic:2002wh}.
Thus, modest hierarchies of these areas can lead to exponential hierarchies in the Yukawa couplings.\footnote{A roughly uniform distribution of such areas could lead to the approximately scale-invariant spectrum of fermion masses, consistent with observations~\cite{Donoghue:1997rn,Gato-Rivera:2014afa}.}
There may also be selection rules such that some of the couplings are forbidden at the perturbative level by the extra anomalous (or non-anomalous) {\uprm}s, but in some cases very small couplings can emerge from nonperturbative string instanton effects.\footnote{For detailed studies see~\cite{Cvetic:2009yh,*Cvetic:2009ez,*Cvetic:2009ng}.}

Of course, there are also other possibilities, such as multiple Higgs bosons, the ordinary (field-theoretic) Froggatt-Nielsen mechanism, etc, that are not unique to string theory, and exponentially-small couplings could have other, non-stringy, origins~\cite{Giudice:2016yja}.

We also remark that there are sometimes relations between  Yukawa couplings that are often different from familiar GUT-type relations.

\subsubsection{Nonstandard Neutrino Mass Mechanisms}
It is nontrivial to incorporate conventional bottom-up type models for neutrino mass in simple perturbative string constructions.
For example, even when there is an underlying grand unification symmetry (as in heterotic models), the
GUT symmetry is often broken in the extra dimensions. Moreover, it is difficult or impossible\footnote{The absence of the {\bf 126} of $SO(10)$ is a theorem
for free-field heterotic constructions~\cite{Dienes:1996yh}.} to obtain the large representations, such as the {\bf 126} of $SO(10)$ that are often employed in model building  to generate large Majorana mass terms such as $W\sim S_{126} N^c N^c$ [suppressing $SO(10)$ indices] for the right-handed (sterile) neutrinos $N^c$, or the non-supersymmetric equivalent. More likely are that
both Majorana and Dirac mass terms that can lead to a seesaw model for the active neutrinos $N$ are generated by stringy higher-dimensional operators analogous to
\refl{uhdo},
\beq 
W=\lambda_D  \frac{1}{M_s^p} L H_u N^c S_1 S_2 \cdots S_p + \lambda_S  \frac{1}{M_s^{q+1}} N^c N^c S_1 S_2 \cdots S_q. \eeql{numass}
However, the additional gauge symmetries 
 sometimes restrict such operators so that the resultant seesaw masses for the active neutrinos are too small to be relevant~\cite{Giedt:2005vx}.
There are a few successful examples~\cite{Lebedev:2007hv} involving rather large $p$ and $q$, but these involve
of $\mathcal{O}$(100) right-handed neutrinos $N^c$ and are therefore quite different from the conventional seesaw.

Type IIA constructions are even more difficult, because they conserve lepton number at the perturbative level~\cite{Ibanez:2001nd,Antoniadis:2002qm}.
However, $L$ can be violated by nonperturbative string instantons~\cite{Blumenhagen:2006xt,Ibanez:2006da}, possibly allowing a Majorana mass for the $N^c$. A direct stringy Weinberg operator $\lambda_T N H_u N H_u/M_s$~\cite{Weinberg:1980bf}
is another possibility~\cite{Cvetic:2010mm}, although there would have to be a low string scale  for it to be sufficiently large,
i.e., $M_s/\lambda_T \sim 10^{14}$ GeV. Alternatively, exponentially-small Dirac masses  can be generated  at the perturbative level if there are large intersection areas $A_I$~\cite{Blumenhagen:2005mu,Blumenhagen:2006ci}, or nonperturbatively by string instantons~\cite{Cvetic:2008hi}.

There are other (not necessarily stringy) possibilities for small Dirac or Majorana masses (or even mixing between active and
sterile neutrinos) from higher-dimension field-theoretic or stringy operators in the superpotential or K\"{a}hler potential~\cite{Langacker:1998ut,ArkaniHamed:2000bq},
non-holomorphic soft supersymmetry breaking terms~\cite{Demir:2007dt}, or large dimensions. The entire subject is reviewed in~\cite{Langacker:2011bi}.

\subsubsection{Perturbative Global Symmetries from Anomalous \uprm}

Exact global symmetries are believed to be incompatible with string theory or quantum gravity~\cite{Banks:1988yz,*Witten:2000dt,*Burgess:2008ri,*Banks:2010zn,*Witten:2017hdv}. However, as has already been described in Section~\ref{zprime}, constructions often involve
anomalous \uprm\ factors. The associated \zpr\ gauge bosons typically acquire string-scale masses, but the symmetry survives as a global symmetry on the perturbative couplings. Such global \uprm\,s are only approximate, and can be broken by non-perturbative string instanton effects~\cite{Blumenhagen:2006xt,Ibanez:2006da,Blumenhagen:2009qh,*Cvetic:2011vz}, which may, however, be very small. Possible implications include Yukawa hierarchies~\cite{Cvetic:2009yh,*Cvetic:2009ez,*Cvetic:2009ng}, suppressed Majorana neutrino masses~\cite{Langacker:2011bi}, and the $\mu$ problem.

\subsection{Additional Issues}

\subsubsection{Grand Unification  and  Gauge Unification}

Most string compactifications do not yield canonical grand unified theories in the low energy four-dimensional theory.
Although heterotic constructions involve an underlying grand unification that can lead
to canonical grand unification via a special choice of gauge field background (as encoded
in a holomorphic vector bundle), the symmetry may be broken in the compactification,\footnote{Orbifold GUTs~\cite{Kawamura:2000ev,*Hall:2002ea,*Raby:2008gh}, involving extra dimensions, model some of the relevant physics.} leading to a \emph{direct compactification} to the standard model, MSSM, or extended gauge group.  Direct compactifications may have some advantages or disadvantages compared to traditional GUTs, such as non-standard Yukawa relations (with obvious implications for the fermion spectrum) or the absence of the doublet-triplet problem. 

Even in the case of a full grand unified gauge group in four dimensions there may not exist the adjoint or other Higgs representations needed to break the GUT symmetry. 
They are also unlikely to yield very large groups or large representations, such as the 126-dimensional Higgs representation
often invoked in $SO(10)$ models of neutrino mass~\cite{Dienes:1996yh}.

Weakly coupled type I and II brane constructions generally do not
have an underlying grand unification\footnote{See~\cite{Gato-Rivera:2014afa} for an interesting discussion of the emergence of certain GUT-type features in brane models.} except for special cases, such as a full $U(5)$ stack of  branes in Type IIA (with, however, a perturbative $t$-quark Yukawa forbidden by the $U(1)$). $SO(10)$ GUTs cannot be achieved at weak coupling due to the
absence of the $16$-dimensional representation. 

Both $SU(5)$ and $SO(10)$ GUTs can be achieved
in F-theory, however, due to strongly coupled effects \cite{Beasley:2008dc,Donagi:2008ca}. The
phenomenology of these models has been studied extensively; see \cite{Heckman:2010bq} and references therein.
Interestingly, they may only arise on subloci in complex structure moduli space \cite{Grassi:2014zxa},
which in simple geometries occur at high codimension \cite{Braun:2014xka} but at lower codimension
in more typical geometries \cite{Halverson:2016tve}. On the other hand, $SU(3)$ and $SU(2)$ are the only $SU(N)$
groups that arise geometrically in F-theory at generic points in complex structure moduli space \cite{Grassi:2014zxa}, suggesting direct compactification to the standard model rather than a GUT.

One consequence of the absence of canonical grand unification is that
string constructions do not necessarily preserve the simple and successful gauge unification
of the MSSM~\cite{Amaldi:1991cn,*Ellis:1990zq,*Giunti:1991ta,*Langacker:1991an}. 
In heterotic constructions the usual \stf-type boundary condition\footnote{ $\alpha_Y\equiv g'^{2}/4\pi$.} at the unification scale  $M_X$,
\beq
\frac{1}{k_Y}\frac{1}{\alpha_Y} =\frac{1}{k_2}\frac{1}{\alpha_2}=\frac{1}{k_3}\frac{1}{\alpha_3}, \text{ with } (k_Y,k_2,k_3)=(\frac{5}{3},1,1),
\eeql{gutbc}
 may be modified by higher Ka\v{c}-Moody levels~\cite{Dienes:1996du}.
For the most common case of canonical \st\ and \sth\ embeddings, \refl{gutbc} is modified 
to allow an arbitrary positive
$k_Y$, depending on the hypercharge embedding.
Of course, $k_Y\ne 5/3$ can in principle be compensated by other effects, e.g., the absence of low-scale
supersymmetry~\cite{Barger:2005gn}. Another stringy effect is  that $M_X$ is expected to be close to the string scale, modulo
threshold effects (which can be large due to string excitations).
Effects such as exotic particles, lack of supersymmetry, multiple
thresholds, and low and (conventional) high-scale threshold corrections, which can also be present in non-string theories,
are also possible.

Similarly, most brane construction do not predict the normalization in  \refl{gutbc}.
For example, in the intersecting
brane constructions the gauge normalization depends on the volumes of the three-cycles wrapped by the stack,
and so there need be no direct relation between the different gauge factors (unless there is a full $U(5)$ stack).
In some special cases~\cite{Gato-Rivera:2014afa,Ibanez:1998rf,Blumenhagen:2003jy,*Blumenhagen:2008aw}, however, there is a relation
\beq
\frac{1}{\alpha_Y} =\frac{1}{\alpha_2}+ \frac{2}{3}\frac{1}{\alpha_3},
\eeql{specialstack}
which is weaker than but compatible with \refl{gutbc}.

\subsubsection{Low String Scale}
There have many suggestions  that the fundamental scale $M_F$ of nature could be  much smaller than the Planck scale $M_P$, perhaps as small as $1-100$ TeV,
therefore solving or alleviating the electroweak hierarchy problem. This could occur if the relation between $M_F$ and gravity is modified by
the existence of additional space dimensions that are  large (e.g.,~\cite{Antoniadis:1990ew,*Antoniadis:1992fh,*Antoniadis:1997zg,*ArkaniHamed:1998rs,*Antoniadis:1998ig,Dienes:1998vh}) compared to the simple dimensional estimate  $L\sim  \mathcal{O}(M_F^{-1})$,  or warped~\cite{Randall:1999ee,*Randall:1999vf}, although
the former case introduces a new hierarchy problem of why $L$ is so large. In the string context, this corresponds to a string scale $M_s \ll M_P$ with some of the extra dimensions large~\cite{Lykken:1996fj,*Shiu:1998pa}. Although such vacua may be rare in the string landscape, they would not only be relevant  to the hierarchy problem, but could even yield observable signatures at the LHC or other colliders if $M_s$ were of $\mathcal{O}$(TeV).
These could include $Z'\,$s from anomalous $U(1)'\,$s~\cite{Berenstein:2006pk,*Berenstein:2008xg,*Anchordoqui:2011eg}, TeV-scale black holes~\cite{Giddings:2001bu,*Dimopoulos:2001hw},
and various stringy excitations~\cite{Antoniadis:1994yi,Friess:2002cc,*Burikham:2004su,*Dong:2010jt,Lust:2008qc,*Anchordoqui:2008di,*Anchordoqui:2009mm,*Lust:2013koa,*Anchordoqui:2014wha,Bianchi:2010es}, including Kaluza-Klein states, winding modes, and Regge trajectories of higher-spin states.\footnote{Higher-spin states could possibly have cosmological consequences even for a large string scale~\cite{Arkani-Hamed:2015bza}.} For a general review, see~\cite{Berenstein:2014wva}.

\subsubsection{Environmental Selection}

An enormous landscape of vacua would be a natural home for ideas of the multiverse and environmental selection~\cite{Barrow:109141,Cahn:1996ag,Tegmark:1997qn,Hogan:1999wh,Susskind:2003kw,Wilczek:2004cr,Livio1022,Weinberg:2005fh,Schellekens:2006xz,*Schellekens:2008kg,Clavelli:2006di,Carr:1070494,Ubaldi:2008nf,Nomura:2012nt,Schellekens:2013bpa,Meissner:2014pma,doi:10.1093/astrogeo/atu077,Gato-Rivera:2014afa,Dienes:2015xua,Schellekens:2015cia,*Schellekens:2015zua,Linde:2015edk,Donoghue:2016tjk,Baer:2016lpj,*Baer:2017cck}.
In particular,  some aspects of our observed world could be determined by anthropic considerations.
Anthropic reasoning is probably the most plausible explanation that has been advanced for a tiny but
non-zero cosmological constant~\cite{Weinberg:1987dv,*Weinberg:1988cp,Bousso:2000xa,*Polchinski:2006gy,*Bousso:2007gp,*Bousso:2012dk}, and could also resolve other fine-tunings such as the smallness
of the electroweak scale and values of the proton, neutron,  electron, and neutrino masses~\cite{Agrawal:1997gf,*Agrawal:1998xa,*Donoghue:2005cf,*Donoghue:2007zz,*Donoghue:2009me,Tegmark:2003ug,*Pogosian:2004hd,Hall:2014dfa}. It could also
explain discrete choices, such as the existence of precisely three large space dimensions (necessary
for stable planetary orbits~\cite{ehrenfest1917way}).  It should also be emphasized that our observed physics need not be anthropically unique: there could well be many other anthropically-suitable vacua~\cite{Harnik:2006vj,Gedalia:2010iy}.
Furthermore, anthropic considerations would not necessarily have to explain
everything: some aspects (e.g., the precise values for the heavy quark masses) could well be essentially random,
while others (e.g., the smallness or vanishing of the strong $CP$ phase) could have conventional dynamical or symmetry explanations.

The multiverse concept requires not only the existence of an enormous landscape, but also that many of the vacua
are de Sitter and metastable. Some aspects related to metastability of the anti D3-brane
de Sitter uplift are still debated (see, e.g.,~\cite{Michel:2014lva,*Polchinski:2015bea,Kachru:2003aw,Intriligator:2006dd,*Intriligator:2007py} and~\cite{Banks:2003es,*Banks:2012hx,Bena:2012vz,*Bena:2014jaa,*Bena:2015kia} for arguments pro and con, respectively,
\jh{and \cite{Sethi:2017phn} for a discussion on flux compactifications giving rise to rolling
solutions}),
but it is a technical issue that should eventually be resolved. Furthermore, there are other
uplift mechanisms other than from anti D3-branes. 

Another requirement is that there should be a method
for sampling the different vacua. Some form of eternal inflation seems ideally suited. At present, inflationary ideas
are supported by most cosmologists and observations~\cite{Guth:2007ng,Guth:2013sya,Linde:2014nna}, but see~\cite{Ijjas:2013vea} for a contrary point of view.
One important prediction of eternal inflation is that our Universe should almost certainly be either  flat or have a small
negative curvature~\cite{Linde:2003hc,Freivogel:2005vv,*Kleban:2012ph,Guth:2012ww}.

Like string theory itself, the ideas of the multiverse and environmental selection are controversial and difficult to test experimentally.  Our view, however, is that the combination of the existence
of the string landscape and the possibility of eternal inflation, together with its solution
to the cosmological constant problem and the fact that it is a quantum theory of gravity,
are sufficiently attractive that one must consider the multiverse and environmental
selection very seriously.
 It is important to pursue every direct  or
indirect experimental probe and theoretical exploration of these ideas.

On the other hand, even if environmental selection does play a role in accounting for certain
aspects of observed physics, other aspects may not admit such an explanation. In these circumstances, mechanistic or remnant interpretations may play an important role. 

\subsubsection{Other Possible Remnants/Effects}
We briefly mention a few other possibilities.
\begin{itemize}
\item The intrinsic nonlocality of string theory opens up the possibility of spontaneous
Lorentz and $CPT$ violation (e.g.,~\cite{Kostelecky:2000mm}), with implications for the propagation, decays, 
and ($\nu$) oscillations of 
high energy $\gamma$'s, $e^\pm$'s, $\nu$'s, and gravity waves~\cite{Kostelecky:2008ts}, and for alternative versions of baryogenesis
 (e.g.,~\cite{Mavromatos:2017cxr}).
 
 \item The possible  time or space variation of moduli could in principle lead to observable  variations of the fundamental ``constants''  of nature (e.g.,~\cite{Uzan:2010pm,Martins:2017qxd}).
 
 \item Perturbative string theory gives a simple explanation~\cite{Font:2013hia} of the apparent absence of massless continuous spin
 representations of the Poincar\'{e} group~\cite{Wigner:1939cj}.

\end{itemize}

\section{Discussion}\label{discussion}
 String theory is an extremely promising and attractive candidate for a consistent quantum-mechanical unification of gravity and the other interactions. Due to the enormous landscape of vacua and the likely high string scale, it will probably not be possible to directly confirm or falsify it in detail. However, 
 in our view the advantages are so large as  to motivate every type of exploration of possible evidence for or against the basic ideas. One possibility involves theoretical progress, perhaps finding a ``smoking gun'', or finding
 a particular vacuum that corresponds in great detail to the observed standard model or MSSM. However,  so far there is no hint of a smoking gun. Furthermore,  given the enormity of the landscape it is unlikely, in the absence of a theoretical breakthrough, that the exact vacuum 
 corresponding to our world will be found.
 
 Another direction, explored in this article, involves the possibility that the standard models
 of particle physics or cosmology are not the final story, and that some types of additional physics are much more likely  than others in the string landscape, especially if one
 restricts to vacua in which  the low energy SM and cosmological degrees of freedom are essentially the ones that emerge
 directly at the string scale and not as composites.
 In particular,
  many semi-realistic constructions contain remnant physics, by which we mean new particles, interactions, or features that are present due to the details of the compactification.  Certain types of remnants (e.g., axions, moduli, large non-abelian gauge sectors, $Z'$ s, quasi-chiral exotics, extended Higgs/Higgsino sectors), occur very frequently. While not unique to string theory, their observation would be at least suggestive. This would especially be the case if they are  unrelated to any standard model problem and are not part of a more complete low-energy theory. In such a scenario they would appear ad hoc to a bottom-up
  model builder, who would likely not add them to their theory. Conversely,
 features such as large representations are extremely rare in the string landscape, and their observation would constitute evidence against string theory. For any feature that is firmly
 determined to be in the swampland, its observation would constitute falsification of string theory.
 
 We conclude with a parable.
Many string vacua look nothing like our world, they are darts that missed the dartboard entirely. Other darts get somewhat closer and hit the board, while still others may be scattered around a bullseye created by theorists that represent the  ``simplest'' or ``most natural'' models. But Nature is not beholden to the
theorists' bullseye. The true bullseye is determined experimentally, and until we do so we 
must take near misses seriously, particularly if they typically have one (or more) of a few
simple features. Hence, remnants.

\vspace{1cm}
\noindent
\textbf{Acknowledgments.} J.H. thanks C. Long for useful discussions. He also thanks the students at TASI 2017
for many interesting discussions, and
Mirjam Cveti{\v c}, Tom DeGrand, and Igor
Klebanov for organizing such an excellent school.
J.H. is supported by
NSF grant PHY-1620526. 

\appendix

\section{String Theory Calculations and Formalism}

The goal of the rest of this review has been to discuss
broadly what is currently known about string remnants, citing
the original literature and presenting the material
in a way accessible to any theoretical high energy physicist.

In this appendix we present some additional details for the interested reader.

\subsection{Dark Gauge Sectors in F-theory}
\label{sec:dark gauge Ftheory}

In this section we wish to show that large gauge sectors in F-theory are completely
generic, due to the prevalence of so-called Non-Higgsable clusters, and furthermore
that this almost always arises at strong coupling. 
We will focus
entirely on F-theory geometry, which sets the foundation on which all F-theory
compactifications are built. See, e.g., \cite{Grassi:2014zxa} for a discussion of
how fluxes may modify aspects of the physics that are determined geometrically.

\subsubsection{F-theory Geometry}
\begin{table}[t]
\begin{center}
\scalebox{.9}{\begin{tabular}{|c|c|c|c|c|c|c|c|c}
\hline
$F$ & $l$ & $m$ & $n$ & Sing. & $G$ & $\tau$ & $g_s$ \\  \hline
$I_0$&$\geq $ 0 & $\geq $ 0 & 0 & none & none & $\mathbb{H}$& $\geq 0$ \\
$I_n$ &0 & 0 & $n \geq 2$ & $A_{n-1}$ & $\gsu(n)$ or $\gsp(\lfloor
n/2\rfloor)$ & $i\infty$ & $0$\\
$II$ & $\geq 1$ & 1 & 2 & none & none & $e^{2\pi i/3}$ & $2/\sqrt{3}$\\
$III$ &1 & $\geq 2$ &3 & $A_1$ & $\gsu(2)$ & $i$ & $1$\\
$IV$ & $\geq 2$ & 2 & 4 & $A_2$ & $\gsu(3)$  or $\gsu(2)$ & $e^{2\pi i/3}$ & $2/\sqrt{3}$\\
$I_0^*$&
$\geq 2$ & $\geq 3$ & $6$ &$D_{4}$ & $\gso(8)$ or $\gso(7)$ or $\gg_2$& $\mathbb{H}$& $\geq 0$ \\
$I_n^*$&
2 & 3 & $n \geq 7$ & $D_{n -2}$ & $\gso(2n-4)$  or $\gso(2n -5)$ & $i\infty$ & $0$\\
$IV^*$& $\geq 3$ & 4 & 8 & $E_6$ & $\ge_6$  or $\gf_4$  & $e^{2\pi i/3}$ & $2/\sqrt{3}$\\
$III^*$&3 & $\geq 5$ & 9 & $E_7$ & $\ge_7$ & $i$ & $1$\\
$II^*$& $\geq 4$ & 5 & 10 & $E_8$ & $\ge_8$  & $e^{2\pi i/3}$ & $2/\sqrt{3}$ \\ \hline
\end{tabular}}
\caption{Kodaira fiber $F_i$, singularity, and gauge group $G_i$ on
the seven-brane at $x_i=0$ for given $l_i$, $m_i$, and $n_i$. In the second last two columns we display the minimal $g_s$ with corresponding $\tau$.}
\label{tab:gauge}
\end{center}
\end{table}

F-theory \cite{Vafa:1996xn,Morrison:1996pp,Morrison:1996na} is a generalization of Type IIB string theory
that allows for strong coupling regions in the extra dimensions of
space $B$, as encoded in a holomorphically varying axiodilaton field
\begin{equation}
\tau = C_0 + i e^{-\phi}.
\end{equation}
One central observation of \cite{Vafa:1996xn} is that it is convenient to encode
this field profile in the complex structure of an elliptic curve
that is fibered over $B$. The Calabi-Yau manifold $X$ of the F-theory
compactification is therefore an elliptic fibration
\begin{equation}
E\hookrightarrow X \xrightarrow{\pi} B.
\end{equation}
So-called birational transformations do not change the complex structure of the smooth
elliptic fibers, and therefore the axiodilaton is invariant under such
transformations. By a theorem of Nakayama, all elliptic fibrations with
a section are birational to a Weierstrass model, which is defined by a hypersurface equation
\begin{equation}
y^2 = x^3 + f x + g, 
\end{equation}
where $f\in \Gamma(-4K), g\in \Gamma(-6K)$ are holomorphic global sections
of of the line bundles $-4K$ and $-6K$ on $B$, respectively. In simple cases where
$B$ is a toric variety there are homogeneous coordinates on the space and
$f$ and $g$ are polynomials functions of appropriate degree in those homogeneous
coordinates.

Above a generic point $p\in B$, $\pi^{-1}(p)=:E_p$ is a smooth elliptic curve. 
On the discriminant locus
\begin{equation}
\Delta=4f^3+27g^2=0,
\end{equation}
points $p\in \{\Delta=0\}$ have $\pi^{-1}(p)$ singular. This can be seen directly from
the cubic 
\begin{equation}
v(x) = x^3 + f x + g.
\end{equation}
Consider a point $p$ away from $\Delta = 0$, the cubic has three distinct roots and the fiber is a double cover
of $\bC$ branched at the roots of $v(x)$ and also a point at infinity. Such a double cover
is a torus, where a basis of one-cycles may be defined by connecting the roots of the cubic
and considering that interval to be the projection of a one-cycle in the double cover. Visualizing
this, suppose that the roots are $p_1$ and $p_2$, and one travels along the bottom sheet from $p_1$
to $p_2$, and then travels back from $p_2$ to $p_1$ along the top-sheet; this defines an $S^1$.
Now consider any path from $p$ to a point inside $\Delta=0$. By definition
$v(x)$ has a degenerate root, signaling that a circle of the type just constructed has collapsed
and the corresponding fiber is singular.  

From the point of view of string theory, the collapsed
fiber gives rise to a monodromy 
\begin{equation}
M =\begin{pmatrix}a&b\\c&d\end{pmatrix} \in SL(2,\bZ)
\end{equation}
associated to the fiber, which means that taking a circle around $\Delta=0$ induces
a map on
\begin{equation}
\tau \mapsto \frac{a \tau + b}{c \tau + d}
\end{equation}
on the axiodilaton. Such a monodromy signals the presence of an object magnetically
charged by the axiodilaton, that is, a seven-brane. In weakly coupled Type IIB compactifications
these may only be D7-branes and O7-planes.

In F-theory there is a richer spectrum
of seven-branes than in weakly coupled Type IIB. They are classified by Kodaira's classification
of codimension one singular fibers, which are determined by the order of vanishing (OOV) of $f$, $g$, and $\Delta$, labeled respectively by $l$, $m$, and $n$ in Table \ref{tab:gauge}. In each
case the OOV determines an $ADE$ singularity in the elliptic fibration and, depending on 
the details of codimension two structure, the gauge algebra on the seven-brane. Unlike
weakly coupled  Type IIB, seven-branes in F-theory support the exceptional simple Lie gauge
groups $G_2$, $F_4$, $E_6$, $E_7$, $E_8$. One can see that these are not weakly coupled
by the study of the $j$-invariant of the elliptic curve. It is given by
\begin{equation}
j=\frac{4f^3}{4f^3+27g^2},
\end{equation}
and in the cases of the exceptional algebras $f$ and $g$ both have non-zero orders of vanishing,
forcing $j$ to be either $0$ or $1$. The $j$-invariant is a function of $\tau$, and in the case
$j=0$ ($j=1$) may be inverted to give $\tau = exp(2\pi i /3)$ ($\tau=i$), up to an $SL(2,\bZ)$
transformation necessary to map $\tau$ into the fundamental domain. These values of $\tau$
have $g_s=2/\sqrt{3}$ and $g_s=1$, respectively; the exceptional seven-branes are strongly
coupled. For a recent in-depth discussion of strongly coupled effects, particularly in
the presence of the elliptic fibers that may arise for non-Higgsable clusters, see \cite{Halverson:2016vwx}.

\subsubsection{Non-Higgsable Clusters}
In recent years it has been discovered that for nearly all toric extra dimensional
spaces $B$, the most general form of $f$ and $g$ appears as
\begin{equation}
f = F \prod_i x_i^{l_i} \qquad g = G \prod_i x_i^{m_i},
\end{equation}
where $F$ and $G$ are sections of appropriate bundles, $x_i$ are homogeneous coordinates
on the toric variety\footnote{For an introduction to toric geometry, see, e.g., \cite{Bouchard:2007ik}. Here, it 
is critical to know that toric varieties are generalizations of weighted projective spaces
that may have many scaling relations for the homogeneous coordinates, rather than one.
Like weighted projective spaces, for any given scaling relation the different homogeneous
coordinates may have different scaling weights.}, and $l_i$ and $m_i$ are positive. Similar statements can be made for
bases that aren't toric (see, e.g., \cite{Morrison:2014lca}), but it is simplest here to discuss the
toric case. When $f$ and $g$ have this form, the discriminant may be written
\begin{equation}
\Delta = \tilde \Delta \prod_i x_i^{n_i},
\end{equation}
where $n_i=min(3 l_i,2 m_i)$. From Table \ref{tab:gauge} it is easy to see that there are non-trivial
Kodaira fibers above each locus $x_i=0$ appearing in the discriminant, and therefore these loci
are also the locations of seven-branes.

Recall from your string theory course that the relative positions of D-branes determines the mass
of the W-boson arising from the string stretched between them. Separating two stacks of D-branes therefore
corresponds to a non-abelian Higgs mechanism. In the current picture, however, there are non-trivial
seven-branes on $x_i=0$ for the most general values of the coefficients of $f$ and $g$, i.e., at
generic points in the complex structure moduli space $\cM_{cs}(X)$ of the elliptic fibration. Therefore
the branes cannot be split at all, and there is no Higgs mechanism! 

\vspace{.5cm}
Such seven-branes are 
called \emph{geometrically non-Higgsable seven-branes} (NH7) and a collection of intersecting NH7's
is called a \emph{geometrically non-Higgsable cluster}. The modifier ``geometrically'' arises is
unnecessary in six-dimensional F-theory compactifications, where all known Higgs mechanisms
are via complex structure deformation, but it is important in four-dimensional compactifications
and two-dimensional compactifications since there the Higgs mechanism may arise in other ways, such
as T-branes \cite{Cecotti:2010bp}. Henceforth we will drop the modifier and refer to these configurations as
either non-Higgsable seven-branes or non-Higgsable clusters. Similarly, since T-branes and fluxes may
break down the gauge group realized geometrically on the seven-brane, such gauge groups are often
referred to as the \emph{geometric gauge group}. Henceforth we will also drop this modifier, and
it is to be understood that all mentioned gauge groups are the geometric gauge groups prior to
any flux or T-brane breaking.

Non-Higgsable seven-branes are almost always strongly coupled. Since an NH7 arises along $x_i=0$ when $l_i$ and $m_i$ are both positive, we see from the 
form of the $j$-invariant that neither all non-Higgsable seven-branes have
\begin{equation}
\tau = i \qquad \text{or} \qquad \tau = e^{2\pi i /3},
\end{equation}
in which case the string coupling is $O(1)$ on the seven-brane. The lone exception is
$I_0^*$ in the case that $(l_i,m_i)=(2,3)$. Then the factors of $x_i$ drop out of the
$j$-invariant and a weak coupling limit may be taken. Such configurations are seven-branes
that become $4$ D7-branes on top of an O7-plane in the weak coupling limit.

The gauge groups that may arise on non-Higgsable seven-branes are limited. From
Table \ref{tab:gauge}, we see that the possibilities are
\begin{equation}
G \in \{E_8,E_7,E_6,F_4,D_4,B_3,G_3, A_2, A_1\},
\end{equation}
This unusual sequence of Lie groups is known as Deligne's exceptional series, and it also
arises in $4d$ $\cN=2$ SCFTs \cite{Beem:2013sza} as the flavor symmetries of SCFTs whose Higgs branch
is the one-instanton moduli space of the instantons of the flavor symmetry. It is interesting to note that $SU(3)$ and $SU(2)$
are the only $SU(n)$ groups that may exist for generic values of the moduli, which is potentially
of phenomenological interest. For example the entire $SU(3)\times SU(2)$ spectrum of the standard model
may be realized at a single intersection of such seven-branes \cite{Grassi:2014zxa}.

There is a rich literature on non-Higgsable clusters. They are very well
understood in six-dimensional compactifications, beginning with the seminal works \cite{Morrison:1996pp,Morrison:2012np,Morrison:2012js} and
also later studies \cite{Taylor:2015ppa, Morrison:2012js,Taylor:2012dr,Morrison:2014era,Martini:2014iza,Johnson:2014xpa,Taylor:2015isa} across a number of topics. Non-Higgsable clusters
conceptually change the task of moduli stabilization since gauge symmetry can exist at
generic points in moduli space ~\cite{Grassi:2014zxa}, i.e., without tuning to a non-trivial codimension in complex structure moduli space~\cite{Braun:2014xka, Watari:2015ysa, Halverson:2016tve}. They
also cannot geometrically give rise to GUTs, but may allow for a simple realization of the Standard Model in F-theory~\cite{Grassi:2014zxa}. In four dimensions, non-Higgsable clusters
 are particularly interesting ~\cite{Morrison:2014lca} due to realizing new structures that do
 not arise in higher dimensions, such as loops~\cite{Morrison:2012np}. NHCs appear in the geometry that is believed to give rise to the largest number of flux vacua~\cite{Taylor:2015xtz}, as well as in recently developed large ensembles of geometries~\cite{Taylor:2015ppa,Halverson:2017ffz,Taylor:2017yqr}; see also \cite{Halverson:2015jua}. In ~\cite{Carifio:2017bov} it was shown that supervised machine learning
 can be used to make progress in understanding large ensembles of geometries with non-Higgsable
 clusters.

Non-Higgsable clusters can be complex. For example,
\cite{Taylor:2015ppa} presented a Monte Carlo analysis of tens of
millions of bases and associated non-Higgsable
clusters, and found that a typical NHC might look like
the figure displayed in section \ref{sec:extendednonabeliananddm},
which is the non-Higgsable cluster associated to
a particular toric threefold base. Each node 
represents an NH7 with the noted gauge group, and
an edge is drawn between any two NH7 that
intersect. Note the presence
of multiple disconnected gauge sectors, complexity
within the graph itself including both loops and long
chains, and five different simple Lie group factors
that appear. It is the prevalence and complexity
of such structures that motivate dark gauge sectors.

\vspace{.5cm}
It will help with physical intuition to turn to some examples. 

All of our examples
will be in the case that $B$ is a toric variety. 
Suppose that $B$ is a toric $n$-fold, which has real dimension $2n$. Then
$f$ and $g$ may be computed from the polytopes
\begin{equation}
\Delta_f = \{m\in \bZ^n \, | \, m\cdot v_i + 4 \geq 0 \,\, \forall\, i \} \, \qquad \quad
\Delta_g = \{m\in \bZ^n \, | \, m\cdot v_i + 6 \geq 0 \,\, \forall \, i \}\,, 
\label{eqn:dfdg}
\end{equation}
which give rise to collections of monomials via
\begin{equation}\label{eqn:mons}
m_{f} \in \Delta_{f} \mapsto \prod_{i}x_{i}^{m_{f}\cdot v_{i} + 4}\,  \qquad \quad
m_{g} \in \Delta_{g} \mapsto \prod_{i}x_{i}^{m_{g}\cdot v_{i} + 6}.
\end{equation}
From this, the most general form of $f$ and $g$ may be computed, and the structure
of associated non-Higgsable clusters may be studied.

Consider one of the first examples of a non-Higgsable cluster \cite{Morrison:1996pp}.
The base of the elliptic fibration is an algebraic surface, the 
Calabi-Yau is a threefold, and the F-theory compactification has
six non-compact directions. Specifically, the base is $B_2=\bF_3$,
the third Hirzebruch surface. This surface may be modeled with
homogeneous coordinates $x_1,x_2, x_3,x_4$ with two $U(1)$ charges,
\begin{center}
\begin{tabular}{ccc}
& $Q_1$ & $Q_2$ \\ \hline
$x_1$ & $1$ & $0$ \\
$x_2$ & $1$ & $0$ \\
$x_3$ & $3$ & $1$ \\
$x_4$ & $0$ & $1$ \\
\end{tabular}
\end{center}
subject to the additional condition that $x_1=x_2=0$ and $x_3=x_4=0$
are removed from the geometry; in technical terms, the Stanley-Reisner
ideal of the associated toric variety is $SRI=\langle x_1x_2,x_3x_4\rangle.$
The anticanonical class $-K_{\bF_3}$ has degree $(5,2)$ and therefore
$f$ and $g$ are polynomials of degree $(20,8)$ and $(30,12)$, respectively.

The existence or non-existence of non-Higgsable clusters is determined
by studying the most general form of $f$ and $g$. This can be done
systematically by determining all monomials associated with points
in $\Delta_f$ and $\Delta_g$ as described above, but we will instead
be descriptive since it better illustrates the central point. Suppose
there was a monomial $m$ in $f$ with a factor $x_4^0$. Since every monomial
in $f$ must be of degree $(20,8)$, we deduce that $m$ must also have a
factor $x_3^8$. But $Q_1(x_3^8)=24$, which overshoots the $20$ required for
$f$. Since all of the charges are positive, this means that some of the
remaining coordinates must be in the denominator, and therefore they
are not global sections; i.e., this monomial $m$ is not defined everywhere
on $\bF_3$, and therefore $m\notin \Gamma(-4K_{\bF_3}).$ Similar arguments
show that there is no monomial in $f$ of the form $x_4^1$. However,
monomials with $x_4^2$ may appear since this would require $x_3^6$ and
$Q(x_3^6)=18<20.$ This analysis can also be done for $g$, with the
result that
\begin{equation}
f = x_4^2 F\, \qquad g=x_4^2 G,
\end{equation}
where $F$ and $G$ are sections of $\Gamma(-4K_{\bF_3}-2D_4)$ and 
$\Gamma(-6K_{\bF_3}-2D_4)$. The discriminant is
\begin{equation}
\Delta = x_4^4\,(4F^3 x_4^2+27G^2) =: x_4^4 \, \tilde \Delta,
\end{equation}
from which we see that we have a seven-brane with a type $IV$ Kodaira
fiber on $x_4=0$. In some cases monodromy may reduce the gauge group
of a type $IV$ fiber from $SU(3)$ to $SU(2)$, but in this example it
does not occur and $G=SU(3)$. 

This $SU(3)$ seven-brane exists for
generic values of the complex structure moduli, as we have seen
directly in this computation, which is easily understood in field theory.
First, the number of adjoint hypermultiplets is $g(C)$, the genus of the
curve $C$ wrapped by the seven-brane. Here the $C=\{x_4=0\}$ and on
$C$ we have $x_3\neq 0$, which leaves $(x_1,x_2)$ as the remaining
coordinates on $C$, subject to the constraint that $x_1=x_2=0$ does
not exist. This, together with the charges of $(x_1,x_2)$, show that
$C$ is a $\bP^1$ and therefore $g(C)=0$; there are no adjoint 
hypermultiplets. Charged hypermultiplets may also arise at 
intersections of the seven-brane with other non-abelian seven-branes
or with the overall residual seven-brane $\tilde \Delta$. In this example
there are no other non-abelian seven-branes and a short computation
shows that $\tilde \Delta = x_4 = 0$ does not exist in the geometry. In
summary, there are no charged hypermultiplets, the gauge theory
on the seven-brane at $x_4=0$ is pure $G=SU(3)$ super Yang-Mills, and
the absence of matter implied that the gauge theory is non-Higgsable.

\cite{Morrison:1996pp} also studied F-theory compactifications on $\bF_n$ for all
$n=0,\dots, 12$. The charges are
\begin{center}
\begin{tabular}{ccc}
& $Q_1$ & $Q_2$ \\ \hline
$x_1$ & $1$ & $0$ \\
$x_2$ & $1$ & $0$ \\
$x_3$ & $n$ & $1$ \\
$x_4$ & $0$ & $1$ \\
\end{tabular}
\end{center}
and $f$ and $g$ are of degree $(4(n+2),8)$ and $(6(n+2),12)$,
respectively, with the same Stanley-Reisner ideal as before.
There is a non-Higgsable seven-brane on $x_4=0$ $\forall n\geq 3$,
satisfying
\begin{center}
\begin{tabular}{c|cccccccccc}
$n$ & $3$& $4$& $5$& $6$& $7$& $8$& $9$& $10$& $11$& $12$ \\ \hline
$G$ & $SU(3)$& $SO(8)$& $F_4$& $E_6$& $E_7$& $E_7$& $E_8$& $E_8$& $E_8$& $E_8$ \\
\end{tabular}.
\end{center}
The case $n=7$ is the most interesting. Here $\tilde \Delta=x_4=0$ does
exist, signifying the presence of matter, but a detailed analysis shows
the matter is a single half-hypermultiplet in the $\textbf{56}$ of
$E_7$. This representation is pseudoreal, and with only a single
half-hypermultiplet there is no gauge invariant holomorphic combination
of the matter fields, signaling the absence of D-flat directions and
therefore also the presence of non-Higgsability. The rest of the cases are non-Higgsable
due to being pure super Yang-Mills.

The origin of non-Higgsability in these examples turns out to cover
all of the six-dimensional examples. In \cite{Morrison:2012np} the possible building
blocks of $6d$ non-Higgsable clusters were classified, with the result
that gauge sectors are either pure super Yang-Mills, or have so little
matter that there is no D-flat direction. Thus, non-Higgsable clusters
are completely understood from the point of view of gauge theory
in $6d$, as determined by an exact dictionary between the geometry and
the gauge theory. 

In $4d$ and $2d$ compactifications the dictionary and
physical origin of non-Higgsability is not yet completely
understood. One natural possibility is that the matter
spectrum and low energy superpotential together
lift any would-be flat directions, so that the
theory is non-Higgsable. However, the authors know
of no reason that such a perturbative Lagrangian
explanation would give rise to Deligne's exceptional
series, and the story may be further complicated by
features that arise in $4d$ and $2d$ compactifications
that do not arise in $6d$ compactifications, such
as D3-brane tadpole cancellation. Obtaining a full
understanding of geometric non-Higgsability is
therefore an important subject in ongoing research.

The geometries giving rise to $4d$ and $2d$
compactifications are relatively well understood, however,
and for the remainder of our discussion we will study
aspects of large ensembles and how they motivate
dark gauge sectors.

\subsubsection{The Genericity of Non-Higgsable Clusters and the Scarcity of Weak Coupling}

We will now discuss an ensemble of F-theory
bases $B$, where in each case the most general
Weierstrass model over $B$ was studied, including
the associated structure of non-Higgsable clusters. This
ensemble is closely related to two other ensembles that we will mention briefly.

\vspace{.5cm}
In all three ensembles, the bases $B$ are toric
varieties and the ensemble is generated by topological
transitions to other bases known as blowups
from some initial bases. We must therefore review
some properties of blow-ups.
For the general case of blow-ups, see, e.g., \cite{griffiths2011principles}. 

Toric blow-ups in toric threefolds, on the other hand,
may be visualized and encoded combinatorially in the
structure of appropriate subdivisions of intervals
or triangles, which correspond to curve and point
blow-ups, respectively. In both types of blow-ups,
the curve or point is related by a new divisor
that is known as the exceptional divisor, and since
the divisor is a toric divisor it may be encoded
in the structure of a new ray. The associated
topological transition changed $h^{11}$ by $1$,
and the change in $h^{31}$ may be approximated
by compute $\Delta_f$ and $\Delta_g$ on both
sides of the transition.

\begin{figure}
\begin{center} 
\begin{tikzpicture}[scale=3]
 \draw[thick,color=Black] (.7101,.5) --
(-.71,.5);\draw[thick,dash pattern={on 1pt off 1pt},color=ForestGreen] (-.71,.5)--(-.51,1.5)--(0,1)--(.5101,1.5)--(.7101,.5); \fill (0,0) circle (.5mm); \fill (.7101,.5) circle (.5mm); \fill
(-.71,.5) circle (.5mm); \fill (0,1) circle (.5mm); \fill (-.51,1.5) circle
(.5mm); \fill (.5101,1.5) circle (.5mm); \draw (0,0) -- (.7101,.5); \draw (0,0) --
(-.71,.5); \draw (0,0) -- (0,1); \draw (0,0) -- (-.51,1.5); \draw (0,0) --
(.5101,1.5); \node at (-.89,.5) {$v_1$}; \node at (.89,.5) {$v_2$}; \node at
(0,-.21) {$0$}; \node at (0,1.151) {$2$}; \node at (-.62,1.6) {$3$}; \node at
(.62,1.6) {$3$}; 
\end{tikzpicture}
  \caption{\label{fig:smalltree} A height-2 and two height-3 blowups of an edge generated by the vertices $\{v_1, v_2\}$.}
\end{center}
\end{figure}
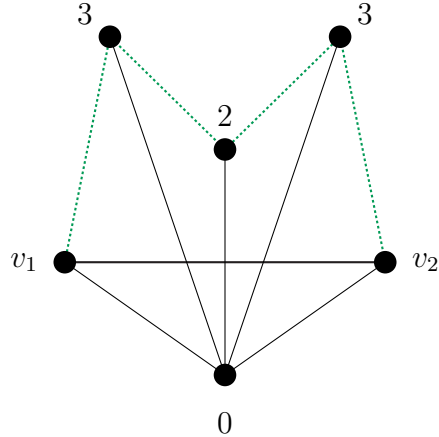

Let us concretely consider the case of a curve blow-up,
given two three-dimensional cones of the associated
toric variety with vertices $(v_0,v_1,v_2)$ and
$(v_1,v_2,v_3)$. To perform the blowup, we add a new
ray $v_e= v_1+v_2$ and replace the original $2$ cones with
$4$ cones generated by 
\begin{equation}
(v_{0}, v_{1}, v_{e}), (v_{0}, v_{2}, v_{e}), (v_{1}, v_{e}, v_{3}), (v_{2}, v_{e}, v_{3}).
\end{equation}
This procedure may be iterated by subdividing the new
cones in the same way, yielding the diagram
shown in Figure \ref{fig:smalltree} after 
two additional blow-ups. Some of the structure in
the figure, such as the line between $v_1$ and $v_2$,
will be discussed when polytopes are introduced.

Point blow-ups proceed in a similar way. Consider a cone
generated by $(v_1,v_2,v_3)$. It may be blown up by
adding the new ray $v_e:= v_1+v_2+v_3$,
where the old cone structure is replaced by
\begin{equation}
(v_1,v_2,v_e) \qquad (v_2,v_3,v_e)\qquad (v_3,v_1,v_e).
\end{equation}
One can visualize this procedure via the
diagram
\begin{center}
\begin{tikzpicture}[scale=1.5, every node/.style={scale=0.9}]
\draw[thick,color=Black] (90:.75) -- (90+120:.75) -- (90+120+120:.75) -- cycle;
\fill (90:.75) circle (.5mm);
\fill (90+120:.75) circle (.5mm);
\fill (90+240:.75) circle (.5mm);
\node at (90:1) {$v_1$}; \node at (90+120:1) {$v_2$}; \node at (90+240:1) {$v_3$};
\draw[thick,->] (1.25,.1) -- (1.75,.1);
\begin{scope}[xshift=3cm]
\draw[thick] (90:.75) -- (0,0);
\draw[thick] (90+120:.75) -- (0,0);
\draw[thick] (90+240:.75) -- (0,0);
\fill (0,0) circle (.5mm);
\draw[thick,color=Black] (90:.75) -- (90+120:.75) -- (90+120+120:.75) -- cycle;
\fill (90:.75) circle (.5mm);
\fill (90+120:.75) circle (.5mm);
\fill (90+240:.75) circle (.5mm);
\node at (90:1) {$v_1$}; \node at (90+120:1) {$v_2$}; \node at (90+240:1) {$v_3$};
\node at (0,-.2) {$v_e$};
\end{scope}
\end{tikzpicture},
\end{center}
where the vertices have been projected into a common
plane in order to make the subdivision structure
clear. The vertices in Figure \ref{fig:smalltree}
could similarly be projected onto the edge between
$v_1$ and $v_2$.

\vspace{1cm} 
Having introduced toric blowups, we now turn to a
discussion of a concrete ensemble. However, we strongly
urge the reader to also study the ensembles presented
in excellent work by Taylor and Wang \cite{Taylor:2015ppa,Taylor:2017yqr}. The ensemble presented
here and the ensembles of Taylor and Wang have
different strengths, but the same qualitative phenomena:
non-Higgsable clusters are completely generic and 
typically arise in multiple disconnected components,
giving rise to large gauge groups and complex dark
sectors.

\vspace{.5cm}
\noindent \emph{An Algorithmic Ensemble of $\frac43 \times 2.96 \times 10^{755}$ Geometries}

We now present the ensemble of \cite{Halverson:2017ffz}. Compared to \cite{Taylor:2017yqr}, it has the
disadvantage of being restrictive in the sense that a certain ``height'' bound is imposed that
is sufficient to have a good F-theory base, but not necessary, and thus is likely significantly
undercounts F-theory geometries.

On the other hand, though
the height bound is restrictive, it is concrete enough that local patches consistent with
the height bound can be classified, giving a concrete construction algorithm for an exponentially
large number of geometries using tree-like structures. Under the assumption that an $O(1)$ number of the bases
in this ``tree'' ensemble
support at least one consistent flux solution, which is rather conservative since a Bousso-Polchinski
argument would imply there are many, this ensemble gives an exact lower bound on
the number of F-theory geometries
\begin{equation}
\text{\# F-theory Geometries} \geq \frac43 \times 2.96 \times 10^{755}.
\end{equation}
Furthermore, by utilizing the construction algorithm, \cite{Halverson:2017ffz} was able to demonstrate
universality in the ensemble. The authors find this avenue promising because --- as is the spirit of remnants --- demonstrating
universality is a natural avenue for making predictions from string theory, and being able to do so
in an exponentially large ensemble is often difficult. In
such cases it may be possible to apply various techniques from
computer science, such as supervised machine learning or genetic algorithms \cite{Abel:2014xta,He:2017dia,Ruehle:2017mzq,Carifio:2017bov}. See section \ref{sec:bigdata}
for further discussion.

\vspace{.5cm}

Let us now describe how the ensemble is constructed.

Consider a smooth weak Fano toric threefold, where
weak Fano means that any holomorphic curve $C$
satisfies $-K\cdot C \geq 0$, where $-K$ is the
anticanonical class. Such a toric threefold
may be encoded in a fine regular star triangulation
of a $3d$ reflexive polytope $\Delta^\circ$, via the
associated fan formed from $3d$ and $2d$ cones
corresponding to faces and edges on the facets of
$\Delta^\circ$. This is an important distinction: the
facets are codimension one faces of $\Delta^\circ$ that
may have interior points that give rise to non-trivial triangulations.
The triangles and lines in that triangulation are also referred to
as faces and edges. Henceforth, ``face'' and ``edge'' means the face
and edge of the triangulation, and we will refer to the codimension
one faces of $\Delta^\circ$ as facets.
For any facet, the number of faces and edges in any of its triangulations is independent 
of the triangulation (as long as all points are used) and is a simple function of the
number of facet vertices and number of lattice points in the facet boundary.

Topological transitions to distinct threefolds arise via smooth toric blowups of curves
or points, as discussed above. Consider a curve associated with a $2d$ cone generated
by $v_1$ and $v_2$. Viewing the facet head on, there would be a corresponding interval
\begin{center}
\begin{tikzpicture}[scale=1.5]
\draw[thick,color=Black] (0,0) -- (1,0);
\fill (0,0) circle (.5mm);
\fill (1,0) circle (.5mm);
\node at (0,.3) {$v_1$};
\node at (1,.3) {$v_2$};
\node at (0,-.3) {$1$};
\node at (1,-.3) {$1$};
\end{tikzpicture}
\end{center}
and the corresponding curve may be blown up. This subdivides the interval  by adding a point on the edge and its corresponding ``height'', giving a sequence of blowups
\begin{center}
\begin{tikzpicture}[scale=1.5]
\draw[thick,color=Black] (0,0) -- (1,0);
\fill (0,0) circle (.5mm);
\fill (1,0) circle (.5mm);
\node at (0,-.3) {$1$};
\node at (1,-.3) {$1$};
\draw[thick,->] (1.25,0) -- (1.75,0);
\draw[thick,dash pattern={on 1pt off 1pt},color=ForestGreen] (2,0) -- (3,0);
\fill (2,0) circle (.5mm);
\node at (2,-.3) {$1$};
\fill (2.5,0) circle (.5mm);
\node at (2.5,-.3) {$2$};
\fill (3,0) circle (.5mm);
\node at (3,-.3) {$1$};
\draw[thick,->] (3.25,.1) -- (3.75,.38);
\draw[thick,->] (3.25,-.1) -- (3.75,-.38);
\draw[thick,->] (5.25,.38) -- (5.75,.1);
\draw[thick,->] (5.25,-.38) -- (5.75,-.1);
\draw[thick,dash pattern={on 1pt off 1pt},color=ForestGreen] (4,.5) -- (5,.5);
\fill (4,.0+.5) circle (.5mm);
\node at (4,-.3+.5) {$1$};
\fill (4.5,0+.5) circle (.5mm);
\node at (4.5,-.3+.5) {$2$};
\fill (4.75,0+.5) circle (.5mm);
\node at (4.75,-.3+.5) {$3$};
\fill (5,0+.5) circle (.5mm);
\node at (5,-.3+.5) {$1$};
\draw[thick,dash pattern={on 1pt off 1pt},color=ForestGreen] (4,-.5) -- (5,-.5);
\fill (4,.0-.5) circle (.5mm);
\node at (4,-.3-.5) {$1$};
\fill (4.5,0-.5) circle (.5mm);
\node at (4.5,-.3-.5) {$2$};
\fill (4.25,0-.5) circle (.5mm);
\node at (4.25,-.3-.5) {$3$};
\fill (5,0-.5) circle (.5mm);
\node at (5,-.3-.5) {$1$};
\draw[thick,->] (5.25,.38) -- (5.75,.1);
\draw[thick,->] (5.25,-.38) -- (5.75,-.1);
\draw[thick,dash pattern={on 1pt off 1pt},color=ForestGreen] (6,0) -- (7,0);
\fill (6,.0) circle (.5mm);
\node at (6,-.3) {$1$};
\fill (6.25,0) circle (.5mm);
\node at (6.25,-.3) {$3$};
\fill (6.5,0) circle (.5mm);
\node at (6.5,-.3) {$2$};
\fill (6.75,0) circle (.5mm);
\node at (6.75,-.3) {$3$};
\fill (7,0) circle (.5mm);
\node at (7,-.3) {$1$};
\end{tikzpicture}
\end{center}
where the edge between $v_1$ and $v_2$ is the edge in a facet, and, for example, the point labeled
by $2$ is the ray $v_1+v_2$ projected into the facet. The sequence of blowups pictured here will be referred
to colloquially as ``trees,'' and accordingly the dashed green lines 
correspond to new edges that arise from
the subdivisions associated to the blowup. Then any additional ray takes the form 
$v_{e_i} = a v_1 + b v_2$
and the combination 
\begin{equation}
h_i = a_i + b_i
\end{equation}
is the ``height" of the associated ray. Since the sequence forms a ``tree," we refer to each ray is as leaf,
and in particular leaves that sit in the original facet are on the ground, and sometimes referred to as
``roots."

Similarly, point blowups can be visualized via the subdivision of one face in the triangulation of a facet.
The first smooth toric blowup would have $v_e=v_1+v_2+v_3$ and the associated diagram is
\begin{center}
\begin{tikzpicture}[scale=1.5, every node/.style={scale=0.9}]
\draw[thick,color=Black] (90:.75) -- (90+120:.75) -- (90+120+120:.75) -- cycle;
\fill (90:.75) circle (.5mm);
\fill (90+120:.75) circle (.5mm);
\fill (90+240:.75) circle (.5mm);
\node at (90:1) {$1$}; \node at (90+120:1) {$1$}; \node at (90+240:1) {$1$};
\draw[thick,->] (1.25,.1) -- (1.75,.1);
\begin{scope}[xshift=3cm]
\draw[thick,dash pattern={on 1pt off 1pt},color=ForestGreen] (90:.75) -- (0,0);
\draw[thick,dash pattern={on 1pt off 1pt},color=ForestGreen] (90+120:.75) -- (0,0);
\draw[thick,dash pattern={on 1pt off 1pt},color=ForestGreen] (90+240:.75) -- (0,0);
\fill (0,0) circle (.5mm);
\draw[thick,color=Black] (90:.75) -- (90+120:.75) -- (90+120+120:.75) -- cycle;
\fill (90:.75) circle (.5mm);
\fill (90+120:.75) circle (.5mm);
\fill (90+240:.75) circle (.5mm);
\node at (90:1) {$1$}; \node at (90+120:1) {$1$}; \node at (90+240:1) {$1$};
\node at (0,-.2) {$3$};
\end{scope}
\end{tikzpicture}
\end{center}
where again the heights are labeled. The collection of trees built on edges (faces)
will be referred to as \emph{edge trees} (\emph{face trees}).

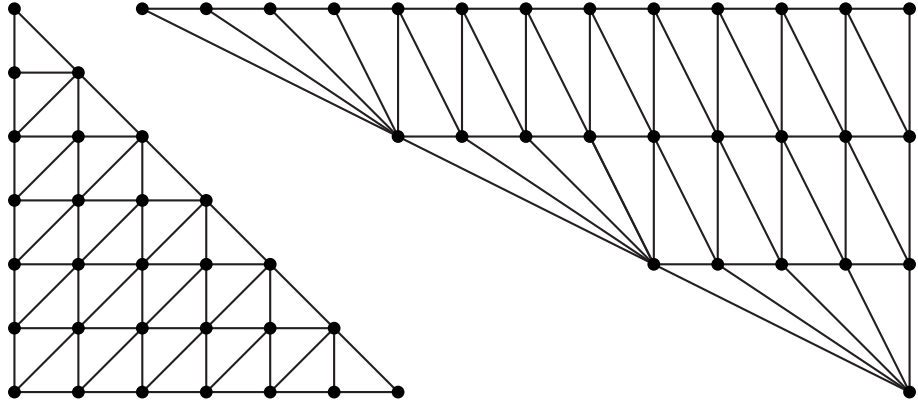
\begin{figure}
\begin{center}
\begin{tikzpicture}[scale=1.7]
\draw[thick,color=Black] (0,0) -- (3,0) -- (0,3) -- cycle;
\draw[thick,color=Black] (0,.5) -- (2.5,.5);
\draw[thick,color=Black] (0,1) -- (2,1);
\draw[thick,color=Black] (0,1.5) -- (1.5,1.5);
\draw[thick,color=Black] (0,2) -- (1,2);
\draw[thick,color=Black] (0,2.5) -- (.5,2.5);
\draw[thick,color=Black] (.5,0) -- (.5,2.5);
\draw[thick,color=Black] (1,0) -- (1,2);
\draw[thick,color=Black] (1.5,0) -- (1.5,1.5);
\draw[thick,color=Black] (2,0) -- (2,1);
\draw[thick,color=Black] (2.5,0) -- (2.5,.5);
\draw[thick,color=Black] (0,2) -- (.5,2.5);
\draw[thick,color=Black] (0,1) -- (1,2);
\draw[thick,color=Black] (0,0) -- (1.5,1.5);
\draw[thick,color=Black] (1,0) -- (2,1);
\draw[thick,color=Black] (2,0) -- (2.5,.5);
\draw[thick,color=Black] (0,1.5) -- (.5,2);
\draw[thick,color=Black] (0,.5) -- (1,1.5);
\draw[thick,color=Black] (.5,0) -- (1.5,1);
\draw[thick,color=Black] (1.5,0) -- (2,.5);
\fill (0,0) circle (.5mm); \fill (0,.5) circle (.5mm); \fill (0,1) circle (.5mm);
\fill (0,1.5) circle (.5mm); \fill (0,2) circle (.5mm); \fill (0,2.5) circle (.5mm);
\fill (0,3) circle (.5mm);
\fill (.5,0) circle (.5mm); \fill (.5,.5) circle (.5mm); \fill (.5,1) circle (.5mm);
\fill (.5,1.5) circle (.5mm); \fill (.5,2) circle (.5mm); \fill (.5,2.5) circle (.5mm);
\fill (1,0) circle (.5mm); \fill (1,.5) circle (.5mm); \fill (1,1) circle (.5mm);
\fill (1,1.5) circle (.5mm); \fill (1,2) circle (.5mm); 
\fill (1.5,0) circle (.5mm); \fill (1.5,.5) circle (.5mm); \fill (1.5,1) circle (.5mm);
\fill (1.5,1.5) circle (.5mm); 
\fill (2,0) circle (.5mm); \fill (2,.5) circle (.5mm); \fill (2,1) circle (.5mm);
\fill (2.5,0) circle (.5mm); \fill (2.5,.5) circle (.5mm);
\fill (3,0) circle (.5mm);
\draw[thick,color=Black] (1,3) -- (7,3) -- (7,0) -- cycle;
\draw[thick,color=Black] (3,2) -- (7,2);
\draw[thick,color=Black] (5,1) -- (7,1);
\draw[thick,color=Black] (6.5,1) -- (6.5,3);
\draw[thick,color=Black] (6,1) -- (6,3);
\draw[thick,color=Black] (5.5,1) -- (5.5,3);
\draw[thick,color=Black] (5,1) -- (5,3);
\draw[thick,color=Black] (4.5,2) -- (4.5,3);
\draw[thick,color=Black] (4,2) -- (4,3);
\draw[thick,color=Black] (3.5,2) -- (3.5,3);
\draw[thick,color=Black] (3,2) -- (3,3);
\draw[thick,color=Black] (7,1) -- (6,3);
\draw[thick,color=Black] (6.5,1) -- (5.5,3);
\draw[thick,color=Black] (6,1) -- (5,3);
\draw[thick,color=Black] (5.5,1) -- (4.5,3);
\draw[thick,color=Black] (5,1) -- (4,3);
\draw[thick,color=Black] (7,2) -- (6.5,3);
\draw[thick,color=Black] (4,2) -- (3.5,3);
\draw[thick,color=Black] (3.5,2) -- (3,3);
\draw[thick,color=Black] (7,0) -- (6.5,1);
\draw[thick,color=Black] (7,0) -- (6,1);
\draw[thick,color=Black] (7,0) -- (5.5,1);
\draw[thick,color=Black] (5,1) -- (4.5,2);
\draw[thick,color=Black] (5,1) -- (4,2);
\draw[thick,color=Black] (5,1) -- (3.5,2);
\draw[thick,color=Black] (3,2) -- (2.5,3);
\draw[thick,color=Black] (3,2) -- (2,3);
\draw[thick,color=Black] (3,2) -- (1.5,3);
\fill (1,3) circle (.5mm); \fill (1.5,3) circle (.5mm); \fill (2,3) circle (.5mm); \fill (2.5,3) circle (.5mm);
\fill (3,3) circle (.5mm); \fill (3.5,3) circle (.5mm); \fill (4,3) circle (.5mm); \fill (4.5,3) circle (.5mm);
\fill (5,3) circle (.5mm); \fill (5.5,3) circle (.5mm); \fill (6,3) circle (.5mm); \fill (6.5,3) circle (.5mm);
\fill (7,3) circle (.5mm);
\fill (7,2) circle (.5mm);
\fill (7,1) circle (.5mm);
\fill (7,0) circle (.5mm);
\fill (6.5,1) circle (.5mm);
\fill (6,1) circle (.5mm);
\fill (5.5,1) circle (.5mm);
\fill (5,1) circle (.5mm);
\fill (6.5,2) circle (.5mm);
\fill (6,2) circle (.5mm);
\fill (5.5,2) circle (.5mm);
\fill (5,2) circle (.5mm);
\fill (4.5,2) circle (.5mm);
\fill (4,2) circle (.5mm);
\fill (3.5,2) circle (.5mm);
\fill (3,2) circle (.5mm);
\end{tikzpicture}
\caption{\label{fig:facets} The largest facets for $\Delta^\circ_1$ and $\Delta^\circ_2$, respectively, each with an arbitrary triangulation.}
\end{center}
\end{figure}

A valid F-theory base must have no divisors with OOV $(4,6,12)$ for $(f,g,\Delta)$; see
the appendix of \cite{Halverson:2017ffz} for a complete discussion of this and related issues.
In this ensemble, restricting any leaf to have height $h\leq 6$ is sufficient to avoid the existence 
of $(4,6)$ divisors, and thus any base built by putting $h\leq 6$ edge trees on edges and $h\leq 6$
face trees on face is a valid F-theory base from the point of view of the $(4,6)$ condition.
Building an ensemble this way then necessitates classifying all such face trees and edge trees.
The number with $h\leq N$ as a function of $N$ is given in Table~\ref{tab:heights}.
\begin{table}
\begin{center}
\begin{tabular}{|c|c|c|}
\hline
$N$ & \# Edge Trees & \# Face Trees \\ \hline
$3$ & $5$ & $2$\\
$4$ & $10$ & $17$\\
$5$ & $50$ & $4231$ \\
$6$ & $82$ & $41,873,645$\\ \hline
\end{tabular}
\caption{\label{tab:heights} The number of possible edge trees and face trees as a function of the maximal height $h$.}
\end{center}
\end{table}

An ensemble $S_{\Delta^\circ}$ may then be built from any $3d$ reflexive polytope $\Delta^{\circ}$ as follows. 
Take a fine regular star triangulation $\mathcal{T}(\Delta^{\circ})$ of $\Delta^{\circ}$, which
defines an initial toric variety. For each edge or face, place an edge tree or face tree with $h\leq 6$ for
each such tree. This defines a large sequence of blowups from the initial toric variety, and the
set of all ways of doing this defines $S_{\Delta^\circ}$. Using 
Table \ref{tab:heights}, the cardinality of the ensemble is
\begin{equation}
|S_{\Delta^{\circ}}| = 82^{\#\tilde{E}\text{ on }\mathcal{T}(\Delta^{\circ})}\times (41,873,645)^{\#\tilde{F}\text{ on }\mathcal{T}(\Delta^{\circ})}\, ,
\end{equation}
where $\#\tilde{E}$ and $\#\tilde{F}$ are the number of total edges and faces on $\mathcal{T}(\Delta^{\circ})$.

By studying all $4319$ reflexive polytopes, the cardinality of each ensemble $S_{\Delta^\circ}$ can be
computed. Two polytopes $\Delta_1^\circ$ and $\Delta_2^\circ$ dominate the ensemble; they are the convex
hulls of the vertex sets
\begin{align*}
S_{1} &= \{ (-1,-1,-1),(-1,-1,5),(-1,5,-1),(1,-1,-1)\}\\
S_{2} &= \{ (-1,-1,-1),(-1,-1,11),(-1,2,-1),(1,-1,-1)\}
\end{align*}
and the cardinalities of the associated ensemble of geometries are
\begin{equation}
|S_{\Delta_{1}^{\circ}}| = \frac{2.96}{3} \times 10^{755} \qquad \qquad |S_{\Delta_{2}^{\circ}}| = 2.96 \times 10^{755}\,.
\end{equation}
The factor of $3$ in the denominator arises from a $\bZ_3$ symmetry of the polytope that extends to a $\bZ_3$ action on the associated toric varieties in the ensemble. These give the exact lower bound
\begin{equation}
\text{\# F-theory Geometries} \geq \frac43 \times 2.96 \times 10^{755}
\end{equation}
that was mentioned above. The number may be enlarged in a number of ways, discussed in \cite{Halverson:2017ffz},
and in particular this is a subset of the ensemble considered in \cite{Taylor:2017yqr}.

  \begin{figure}[t]
\begin{center}
\begin{tikzpicture}[scale=3]
\draw[thick,color=Black] (0,0) -- (3,0) -- (0,3) -- cycle;
\draw[thick,color=Black] (0,.5) -- (2.5,.5);
\draw[thick,color=Black] (0,1) -- (2,1);
\draw[thick,color=Black] (0,1.5) -- (1.5,1.5);
\draw[thick,color=Black] (0,2) -- (1,2);
\draw[thick,color=Black] (0,2.5) -- (.5,2.5);
\draw[thick,color=Black] (.5,0) -- (.5,2.5);
\draw[thick,color=Black] (1,0) -- (1,2);
\draw[thick,color=Black] (1.5,0) -- (1.5,1.5);
\draw[thick,color=Black] (2,0) -- (2,1);
\draw[thick,color=Black] (2.5,0) -- (2.5,.5);
\draw[thick,color=Black] (0,2) -- (.5,2.5);
\draw[thick,color=Black] (0,1) -- (1,2);
\draw[thick,color=Black] (0,0) -- (1.5,1.5);
\draw[thick,color=Black] (1,0) -- (2,1);
\draw[thick,color=Black] (2,0) -- (2.5,.5);
\draw[thick,color=Black] (0,1.5) -- (.5,2);
\draw[thick,color=Black] (0,.5) -- (1,1.5);
\draw[thick,color=Black] (.5,0) -- (1.5,1);
\draw[thick,color=Black] (1.5,0) -- (2,.5);
\fill (0,0) circle (.5mm); \fill (0,.5) circle (.5mm); \fill (0,1) circle (.5mm);
\fill (0,1.5) circle (.5mm); \fill (0,2) circle (.5mm); \fill (0,2.5) circle (.5mm);
\fill (0,3) circle (.5mm);
\fill (.5,0) circle (.5mm); \fill (.5,.5) circle (.5mm); \fill (.5,1) circle (.5mm);
\fill (.5,1.5) circle (.5mm); \fill (.5,2) circle (.5mm); \fill (.5,2.5) circle (.5mm);
\fill (1,0) circle (.5mm); \fill (1,.5) circle (.5mm); \fill (1,1) circle (.5mm);
\fill (1,1.5) circle (.5mm); \fill (1,2) circle (.5mm); 
\fill (1.5,0) circle (.5mm); \fill (1.5,.5) circle (.5mm); \fill (1.5,1) circle (.5mm);
\fill (1.5,1.5) circle (.5mm); 
\fill (2,0) circle (.5mm); \fill (2,.5) circle (.5mm); \fill (2,1) circle (.5mm);
\fill (2.5,0) circle (.5mm); \fill (2.5,.5) circle (.5mm);
\fill (3,0) circle (.5mm);
\node [label={[xshift=-.32cm, yshift=.05cm]$v_{0}$}] at (1,1) {};
\end{tikzpicture}

\caption{The largest facet $F_1$ of the 3d reflexive polytope $\Delta_1^\circ$, with an arbitrary triangulation. A type $II$ fiber above the divisor corresponding to $v_0$ places strong constraints on the existence of the Sen limit.}
\label{fig:bigfacetbigone1}
\end{center}
\end{figure}
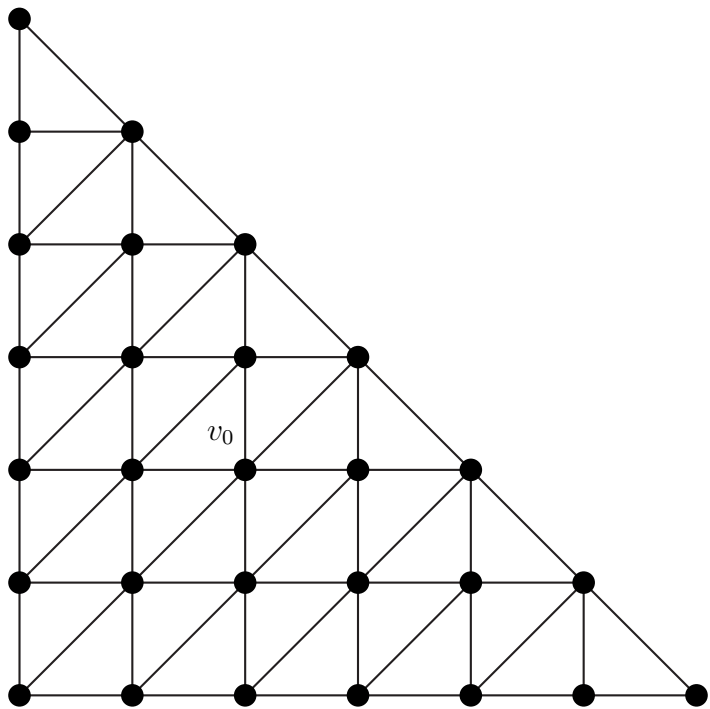

\vspace{.5cm}
A number of universal physical features are known about this ensemble. We refer the interested student
to the original literature for details, and simply cite some of the results. 

In \cite{Halverson:2017ffz}
it was shown that the probability that a base $B$ in the ensemble has a non-Higgsable cluster is 
\begin{equation}
P(\text{NHC}) \geq 1 - 1.07\times 10^{-755},
\end{equation}
so non-Higgsable clusters are completely generic in the ensemble, and understanding why they are implies that
they should be generic in the full set of F-theory geometries. With probability above $0.999995$, the geometric gauge group of the non-Higgsable
clusters is
\begin{equation}
G \geq E_8^{10} \times F_4^{18} \times U^8 \times F_4^{H_2} \times G_2^{H_3} \times A_1^{H_4},
\end{equation}
in which case $rk(G)\geq 160$. $H_i$ is the number of height $i$ leaves above leaves on the ground that carry
$E_8$, and the $\geq$ means that $G$ can enhance beyond this, but this is the minimal structure.
Notably absent are $SU(3)$, $E_6$, $SU(5)$, and $SO(10)$, which are interesting from the perspective of
particle physics. One solution is that flux may break down the geometric gauge group to these factors, or 
they may arise geometrically with some smaller probability. For example, by random sampling the ensemble
it was observed that $E_6$ arises on a distinguished divisor with probability one in a few thousand.

In \cite{Carifio:2017bov} a supervised machine learning algorithm known as logistic regression was used to study those
random samples. With $98\%$ accuracy, the trained algorithm was able to distinguish whether or not the
vertex $v_{E_6}$ associated to a distinguished divisor had an $E_6$ gauge group. By studying how the
machine made its decision, a conjecture was generated that was rigorously proven, so that it can
be computed explicitly that 
\begin{equation}
P(E_6\,\, \text{on}\,\, v_{E_6}\,\, \text{in}\,\, T) \simeq .00059128,
\end{equation}
in which case the number of bases with $E_6$ on that divisor in the ensemble is $5.83 \times 10^{751}$.
Through conjecture generation, a numerical machine learning result was turned into a rigorous result
about the physics of an exponentially large ensemble.

Finally, it is interesting to understand whether or not weakly coupled limits are possible in the ensemble.
Since nearly all geometries have non-Higgsable clusters, and non-Higgsable clusters are intimately tied
to strong string coupling \cite{Halverson:2016vwx}, at generic points in their moduli space the theory is strongly coupled
in $g_s$. Technically, this means that $g_s$ varies over the spacetime, but there are regions in
the extra dimensions where $g_s$ is $O(1)$. However, this does not preclude the possible existence of a
weakly coupled Sen limit. A necessary condition for the existence of a Sen limit is that the non-Higgsable clusters must have singularities
no worse than Kodaira $I_0^*$, and any singularities less severe than $I_0^*$ must enhance to $I_0^*$. Recall
that $I_0^*$ may become $4$ D7 on top of an $O7$ in the weak coupling limit.

In \cite{Halverson:2017vde} it was show that this necessary condition is effectively never satisfied. Specifically,
\begin{equation}
P(\exists \,\, \text{a Sen limit}) \leq 3.0 \times 10^{-391}.
\end{equation}
The authors believe that this result is the first quantitative statement about the size of the
weakly coupled Type IIB string landscape within its larger F-theory context. 

This fact poses a sharp question
for future research: do the qualitative remnants and lessons from weakly coupled Type IIB compactifications
survive the F-theory limit? For some remnants, such as the axions, dark gauge sectors, and standard model exotics, the answer seems to be yes. For example, the existence of large 
dark gauge sectors is strongly motivated by the genericity of non-Higgsable clusters discussed in this section. 

\subsubsection{Dark Gauge Sectors and Glueballs}

The results of the previous sections strongly motivate hidden gauge sectors in string theory. The results were
specific to the broad (relative to Type IIB) context of F-theory, 
but qualitatively match results in other corners of the
landscape, which also regularly have large hidden gauge sectors.
Given what is currently known, it is reasonable to expect that large hidden gauge sectors
are extremely common in the landscape as a whole, with a wide variety of ultraviolet gauge couplings
as determined by moduli stabilization, as well as a variety of matter spectra, which in turn sets
whether the gauge sector is asymptotically free.
Some of the phenomenological considerations related to hidden gauge sectors were discussed 
in the main text.

Here we focus on the possibility that  dark matter comes in a dark confining gauge sector. Whether or
not this arises in a particular construction depends on the beta function and matter spectrum, but with
the plethora of gauge sectors arising in string theory it seems reasonable to expect that some of them
confine.
For simplicity, we will also assume that pure Yang-Mills theory arises in
the infrared, so that it is Yang-Mills that confines. In the presence of matter fields, or alternatively if
the SUSY breaking scale is below the confinement scale, the spectrum of composite dark particles can
become significantly more complex; see, e.g., \cite{Boddy:2014yra}. However, dark pure Yang-Mills will suffice to make some of qualitatively
important points clear, and it is also well motivated.

Consider a scenario \cite{Carlson:1992fn} in which a dark Yang-Mills sector with gauge group $G$ and confinement
scale $\Lambda$ is reheated in the early universe to a temperature $T_{rh}' > \Lambda$. This dark sector 
interacts gravitationally with the standard model and may also have other interactions. It is a thermal
bath of dark gluons and cools as the universe expands. At a temperature $T_\Lambda'\simeq \Lambda$
the gluons confine into glueballs. The effective theory of the glueballs may have $3\to 2$ interactions,
which change the dependence of the dark sector temperature $T'$ on the scale factor $a(t)$, relative
to the scale-factor dependence of the temperature $T$ of a bath of non-interacting non-relativistic particles. 
The temperature ratio is
\begin{equation}
\frac{T'}{T}\propto \frac{a}{ln(a)}.
\end{equation}
The dark sector stays warm because the glueballs ``cannibalize'' themselves, i.e, the $3\to 2$ interactions
increase the average kinetic energy of glueballs while depleting their number. When those interactions cease
to be effective, the glueballs freeze out and leave a relic. The relic abundance of glueballs is approximately
\begin{equation}
\Omega_{gb}h^2 \simeq \frac{\Lambda}{3.6\, \text{eV} \, \xi},
\label{eqn:relic}
\end{equation}
where $\xi:= s/s'$ is the ratio of comoving entropy densities between the visible and dark sector, which remains
constant throughout the process. Its value is set by reheating, and if reheating is not preferentially into
one sector then $\xi=1$.  The approximation here is reasonable because the dependence of the the relic abundance
on other parameter is rather weak; see \cite{Halverson:2016nfq} for an exact expression and a discussion.
We see that in the democratic scenario $\xi=1$ a confinement scale $\Lambda \simeq O(10)\, \text{eV}$, is required to saturate the
relic abundance.

Now consider the fact that string compactifications often have many such gauge sectors.
Each has an ultraviolet gauge coupling $\alpha_{UV}$ that is determined by moduli
stabilization, and accordingly a compactification with many non-abelian sectors that 
confine will have a wide variety of confinement scales. From \eqref{eqn:relic} it is
clear that as the number of confining dark sectors goes up, the likelihood that at least
one will oversaturate the relic abundance increases, since the confinement
scales will have some non-trivial distribution determined by moduli stabilization. 
However, it only takes one oversaturating confining sector to rule out the
model, and therefore it seems that there is a dark glueball problem
in the landscape.

One way out is for the dark matter to decay, but for sectors that only couple to the standard
model via moduli, axions, and gravity the associated decay is usually too slow 
\cite{Halverson:2016nfq}, though in non-thermal cosmologies the problem is partially
alleviated \cite{Acharya:2017szw}. In cases where all of the confining sectors have
sufficiently strong interactions with the visible sector, for example via light mediators,
the problem can also be alleviated; see \cite{Carlson:1992fn,Faraggi:2000pv,Soni:2016gzf,Forestell:2016qhc,Halverson:2016nfq,Dienes:2016vei,Soni:2016yes,daRocha:2017cxu,Acharya:2017szw,Soni:2017nlm,Forestell:2017wov} for a broad dark glueball literature that considers a number of different possible interactions with the standard model.
Another way out of the problem is to make $\xi$ extremely large,
which would constitute an extremely asymmetric reheating process that reheats the
visible sector, but barely reheats the dark sectors. See \cite{Halverson:2016nfq}
for necessary values of $\xi$ under certain circumstances, and \cite{Adshead:2016xxj,Hardy:2017wkr} for recent studies of asymmetric reheating the demonstrate both interesting
possibilities and difficulties.

None of the proposed mechanisms for avoiding the problem are completely compelling in light of a large landscape
with many vacua, and the typicality of many non-abelian gauge sectors. Further alleviating
the problem is an interesting direction for future research, and it would be interesting
to understand how solutions correlate with other remnants.

\subsection{Axions}
\label{app:axions}

Axions are ubiquitous in string compactifications, due
to the many $p$-form gauge fields that may give rise to axions
upon compactification. In this section we review the
general formalism that gives rise to axions in $4d$
compactifications, and will then comment on how they
arise in specific corners of the landscape. We will
also discuss some aspects of axion cosmology, including
mass scales and ultralight axion dark matter.

Consider a $D$-dimensional quantum field theory with $D>4$
and a $p$-form gauge field with field strength
of the form
\begin{equation}
\int d^D x \,\, dC_p \wedge \star dC_p.
\end{equation}
$C_p$ transforms under the gauge symmetry as
\begin{equation}
C_p \mapsto C_p + d\Lambda_{p-1}
\end{equation}
$(p-1)$-form $\Lambda$. $C_p$ may be present in other
interactions such as Chern-Simons terms, but what matters
here is that it is a dynamical higher dimensional field
with a gauge symmetry.

Now consider compactification of this theory to
four dimensions by the ansatz
\begin{equation}
X_D = X_4 \times M,
\end{equation}
where $X_4$ is a non-compact four-manifold and is
a smooth $(D-4)$-manifold. $C_p$ may have terms with
$p$ legs along $M$, in which case we expand in a
basis of non-trivial $p$-forms as
\begin{equation}
C_p = \sum_i a_i(x)\, \omega_i \qquad \qquad \omega_i \in H^p(X,\mathbb{Z}),
\end{equation}
where $x$ is the coordinate on $X_4$. The four-dimensional
fields $a_i(x)$ are axions, and
\begin{equation}
\text{\# axions} = b_p(M),
\end{equation}
where $b_p(M)$ is the $p^\text{th}$ Betti number of $M$.
The axions will appear in the four-dimensional effective
according to the details of the $D$-dimensional theory,
which often includes the interactions of a higher dimensional
supergravity theory in string contexts.

The basic setup occurs essentially everywhere in the landscape.
For example, in different corners the field $C_p$
may be considered to be
\begin{itemize}
\item \textbf{M-theory.} $C_3$, the three-form of $11d$ SUGRA.
\item \textbf{F-theory and IIB.} $B_2$, the NSNS $2$-form. $C_0, C_2$, $C_4$, $B_2$. There are the RR $0$1-form, $2$-form, and $4$-form, and the NSNS $2$-form.
\item \textbf{IIA} $C_1, C_3$, $C_5$, $B_2$. There are the RR $1$1-form, $3$-form, and $5$-form, and the NSNS $2$-form. Note that in case where $M$ is a Calabi-Yau threefold with holonomy
$SU(3)$ (rather than a subgroup) there are no axions from
$C_1$ or $C_5$ since $b_1(M)=b_5(M)=0$.
\item \textbf{Heterotic.} $B_2$, the NSNS $2$-form.
\end{itemize}
The authors are not aware of a well-motivated corner of the
landscape that does not give rise to axions.

All concrete knowledge of string geometries to date
imply that many axions should be expected. Let us discuss
three ensembles to demonstrate the point. In \cite{Kreuzer:2000xy}, Kreuzer
and Skarke classified four-dimensional reflexive polytopes.
A toric variety may be associated in a canonical way to
each fine regular star triangulation of this polytope, and
there is a family of smooth Calabi-Yau threefolds that arise
as hypersurfaces in this toric variety. There are $30,108$
distinct Hodge pairs $(h^{11},h^{21})$ for these threefolds,
and typically both Hodge numbers are in the tens or hundreds.
In recent years there has also been significant progress
in constructing $G_2$ manifolds. These are seven-manifolds
with holonomy $G_2$, which are the relevant manifolds for
$4d$ $\mathcal{N}=1$
compactifications of $11d$ SUGRA or M-theory. Both $G_2$
manifolds and Calabi-Yau threefolds have metrics with
specific holonomy groups, but unlike Calabi-Yau threefolds
there is no simple topological criterion that can be
checked to guarantee the existence of an appropriate
metric; there is no analog of Yau's theorem. However,
in a construction known as the twisted connected sum
construction of $G_2$ manifolds (due to Kovalev \cite{2000math.....12189K},
Corti Haskins Nordstrom Pacini \cite{Corti:2012kd} and others), there
are now at least $50$ million compact $G_2$ manifolds.
In \cite{Corti:2012kd} it was shown that $b_3$ is often of $O(30)$,
but it was also argued convincingly that this ensemble
is only scratching the surface. In the simplest generalizations
which still used so-called ``orthogonal gluing'', all
such $G_2$ manifolds will have $b_3\geq 23$, and the
number could be much higher. See \cite{Halverson:2014tya,Halverson:2015vta,Braun:2016igl,Braun:2017ryx,Braun:2017uku,Braun:2017csz} for physics works on twisted connected
sum $G_2$ manifolds.  Finally, typical geometries
in the previously mentioned F-theory ensembles \cite{Taylor:2015ppa,Halverson:2017ffz,Taylor:2017yqr}
have $h^{11}\simeq O(1000)$, which gives rise to that number
of axions via reduction of the RR $4$-form $C_4$.
In summary, this evidence points to the existence of
at least $O(10)$ axions, and likely $O(100)$ or $O(1000)$.
Henceforth we will refer to ``hundreds" of axions for brevity.
These many-axion scenarios in string theory are sometimes
referred to as an axiverse \cite{Arvanitaki:2009fg}.

Each axion enjoys a perturbative shift symmetry,
the origin of which is the gauge symmetry of $C_p$.
However, the shift symmetry may be broken by
Euclidean instantons that couple to $C_p$ in the 
extra dimensions, an therefore generate non-perturbative
corrections to the $4d$ effective 
action that involve the instantons. For example, non-perturbative
superpotentials of the form
\begin{equation}
W_{np} = A(\phi) e^{-T} + \dots
\end{equation}
may arise, where $a=Im(t)$ and $A(\phi)$ is a holomorphic
function of other superfields in the theory.
Since the Euclidean instanton couples to $C_p$, it 
is a $(p-1)$-brane instanton, which in the respective
corners may be
\begin{itemize}
\item \textbf{M-theory.} $M2$-brane instantons.
\item \textbf{F-theory and IIB.} Worldsheet instantons,
$ED(-1)$-instantons, $ED1$-instantons, and $ED3$-instantons.
\item \textbf{IIA} $ED0$-instantons, $ED2$-instantons,
$ED4$-instantons, and worldsheet instantons.
\item \textbf{Heterotic.} String worldsheet instantons.
\end{itemize}
Each theory with a $B_2$ field also has a dual $B_6$ field
that charges NS5-branes, which may give rise to NS5-brane
instanton corrections.

\subsubsection*{Masses, Axion Cosmology, and Dark Matter}

Axions may have a very wide range of masses in string
theory. In practice it is very hard to actually compute
the masses, since it requires all of the details of 
moduli stabilization. However, in the two most-studied
moduli stabilization scenarios of KKLT \cite{Kachru:2003aw}
and LVS it has been done. A variety of masses are possible; see, e.g., \cite{Stott:2017hvl} and references therein.

We emphasize the importance and difficult of the problem. Computing axion masses in KKLT and LVS, as well as in
more general $\mathcal \cN=1$ compactifications, requires
a number of essential steps.
\begin{itemize}
\item Find all (or all leading, if an approximation can be justified) instanton corrections. For simplicity, consider
instanton
corrections to the superpotential,
\begin{equation}
W_{NP}=\sum_i \mathcal{O}_i e^{-Q_{ij} T_j},
\end{equation} 
where the details of the Q-matrix are determined by the
cycle class, and systematically 
determining the cycle class of contributing
instantons is computationally complex in analogy
to the negative solution to Hilbert's tenth problem \cite{Cvetic:2010ky}.
\item Compute the full scalar potential, including instantons.
	In supersymmetric theories this includes the F-term
	scalar potential
	\begin{equation}
	V_F = e^K \left(K^{I \overline J} D_I W \overline{D_JW}-3|W|^2 \right)
	\end{equation}
	which itself becomes a lengthy polynomial in the hundreds
	of axion fields. Thus, moving from the superpotential
	to the scalar potential is labor intensive for typical
	numbers of axions and instantons.
\item Minimize the potential and search for metastable 
	de Sitter vacua. This is a very difficult problem
	for polynomials of hundreds of variables.
\end{itemize}
For large values of Hodge numbers, which is where most of the
landscape is expected to be located via, e.g., arguments
of Bousso-Polchinski \cite{Bousso:2000xa}, it is difficult
to systematically compute instanton corrections, and
therefore also the scalar potential and its minima. 

Systematic
computation of axion masses, decay constants, and couplings
are critical to furthering our understanding of 
the 
phenomenology and cosmological implications of axions
in the string landscape. However,
recent state-of-the-art work \cite{Braun:2017nhi} has made progress on efficiently
studying certain instanton zero modes for Kreuzer-Skarke
Calabi-Yau threefolds at large $h^{11}$, including the
example with largest $h^{11}$, which has $h^{11}=491$.
These instanton zero modes are important for determining
whether or not there is an instanton correction to the
superpotential, and with additional progress on
other zero modes it may be possible in the next few years
to systematically compute instanton corrections at large 
$h^{11}$ in Type IIB compactifications. This would not solve 
the problem of actually finding the minima, but it is a
necessary first step and a promising direction.

Since it is difficult to do the concrete string calculations
given current mathematical technology, it is worthwhile to
consider effective field theory arguments for axion masses,
where parameters in the effective theory are motivated by
string considerations. We proceed in three steps.
First, determine all symmetries of the EFT. A minimal
	assumption for a realistic EFT with an axion is that
	the symmetries are $G_{SM}=SU(3)\times SU(2)\times U(1)$
	and the shift symmetry of the axion.
Next, determine operators that may give rise to
	axion masses consistent with the symmetries of the
	EFT. For the above minimal possibility, we have
	\begin{equation}
		V_a = A \frac{(h^\dagger h)^n}{\Lambda^{2n-4}} cos\left(\frac{a}{F}\right)
	\end{equation}
	in which case the axion mass
	\begin{equation}
	m_a = \sqrt{A} \, \left(\frac{\Lambda_{EW}}{\Lambda}\right)^{n-1}\, \left(\frac{\Lambda}{F}\right)\, \Lambda_{EW},
	\end{equation}
	where $A$ is a Wilson coefficient, $h$ is the Standard Model
	Higgs field, $F$ is the axion decay constant, 
	$\Lambda$ is the ultraviolet cutoff, and $n$ is
	a non-negative integer whose value is determined
	by the structure of instanton zero modes. 
	The axion mass is a function of the Wilson coefficient,
	the hierarchy between the electroweak scale and the UV cutoff,
	the ratio of the cutoff to the axion decay constant, and the
	electroweak scale itself. In the $n=0$ case that does not involve
	Higgs fields, the electroweak scale drops out, as it should. 

	In string compactifications there is a distribution of axion
	decay constants, see, e.g., \cite{Long:2014fba} and
	often the UV cutoff and mean of the decay constant distribution are an order
	of magnitude or two below the Planck scale. (More specifically, in certain ensembles of well-understood geometries, while the largest decay constant may be near the Planck scale, the mean may be much lower). We therefore
	take
	\begin{equation}
	\Lambda \simeq F \simeq  B M_p,
	\end{equation}
	where $B\sim 1/100 - 1/10$. Then the axion mass is
	\begin{equation}
	m_a \simeq \sqrt{A} \,\, \Lambda_{EW}\,\, \left(\frac{\Lambda_{EW}}{M_{pl}}\right)^{n-1} B^{1-n},
	\end{equation}
	and depends on the Wilson coefficient, the parameter $B$, and electroweak scale, and the electroweak
	hierarchy.
	Under standard effective field theory arguments the Wilson coefficient $A$ is expected to be
	$O(1)$, and therefore the axion mass scale is set by the electroweak scale and hierarchy,
	up to the powers of $B$ that play a minor role relative to the hierarchy. Estimating the
	scales by taking $A=1$, $\Lambda_{EW} = 10^2\,  \text{GeV}$, $\Lambda_{EW}/(BM_{pl}) = 10^{-16}$, we obtain
	axion mass scales
	\begin{align}
	n=0&: \qquad m_a \simeq 10^{27}\, \text{eV}\simeq M_{pl}\nonumber \\
	n=1&: \qquad m_a \simeq 10^{11}\, \text{eV}\simeq \Lambda_{EW} \nonumber \\
	n=2&: \qquad m_a \simeq 10^{-5}\, \text{eV}\simeq m_\nu \nonumber \\
	n=3&: \qquad m_a \simeq 10^{-21}\, \text{eV}\simeq m_{ul} \nonumber \\
	n=4&: \qquad m_a \simeq 10^{-37}\, \text{eV}. 
	\label{eqn:axmasseft}
	\end{align}
We see the axion mass generated by this operator is Planck scale, electroweak scale,
neutrino scale, so-called ``ultralight'' scale, and $10^{-37}\, \text{eV}$ in the
case $n=0,1,2,3,4$, respectively.

This analysis should concern the astute reader: in
the typical case $n=0$ that does not utilize the Higgs field in the axion mass operator,
why is the axion at Planck scale, and how does this square with the notion that axions
should be light? The caveat lies in the assumption $A=1$. In gauge theoretic or string theoretic
setups, $A$ is exponentially suppressed since the operator is generated by an instanton. $A$ is
an instanton suppression factor, and therefore the mass scale conclusions \eqref{eqn:axmasseft}
are correct up to this critical instanton suppression factor.
In string compactifications, the instanton suppression factor is usually determined by the volume
of the cycle wrapped by the instanton in the extra dimensions, which itself is set by the vacuum expectation
value of four-dimensional scalar fields such as K\" ahler moduli in the case of Calabi-Yau threefolds.
In a metastable vacuum, however, these moduli are stabilized and the volumes and instanton suppression
factors are determined; see, e.g., \cite{Cicoli:2013cha}.

To determine axion mass spectra concretely it is therefore critical to know distributions of instanton
suppression factors in metastable vacua. Most metastable vacua are expected to occur in cases where the
number of moduli controlling extra dimensional volumes is large, such as K\" ahler moduli in the case
of Calabi-Yau manifolds and $G_2$ moduli in the case $G_2$ manifolds. This is an important direction
for future research, and until such distributions are determined one can only do analyses at low numbers of
moduli or using EFT arguments with the critical caveat that we have discussed.

\vspace{1cm}
Axions may be cosmologically relevant. For example, depending on details of the model they may be
either the inflaton or the dark matter. We will focus on the possibility of axion dark matter, and refer
the reader to \cite{Baumann:2014nda} and references therein for a detailed treatment of axion inflation and axion monodromy inflation in string compactifications.

Consider evolution of an axion $\phi$ in a Friedmann-Robertson-Walker (FRW)
universe with scale factor $a(t)$. The equation of motion for the axion is
\begin{equation}
\ddot \phi + 3 H \dot \phi + m_\phi^2 \,\phi = 0,
\end{equation}
where $H=\dot a /a$ is the Hubble parameter and $m_\phi$ is the axion mass. If 
$H >> m_\phi$ then Hubble friction freezes $\phi$ and $w_\phi = -1$, where 
$w_\phi = p_\phi / \rho_\phi$ with $p_\phi$ and $\rho_\phi$ the axion pressure
and energy density. On the other hand, if $H<<m_\phi$ then $\phi$ oscillates about
its minimum and behaves like matter (or ``dust"), with an equation of state $w_\phi=0$.

\vspace{.5cm} 
For the axion to be the dark matter in our universe, as opposed to some other matter-like relic,
it must have a number of non-trivial properties. We will focus on two: behaving like matter prior
to matter-radiation equality, and saturating the dark matter relic abundance.

Recall that big bang nucleosynthesis (BBN) makes 
successful predictions for light element
relic abundances. For this to occur, the universe must be radiation dominated at BBN time. On the other
hand, at later times dark matter comes to dominate the energy density of the universe, as is critical
for the formation of large scale structure. It occurs for a simple reason.
Recall that the energy density of a perfect fluid
in an FRW universe scales with $a(t)$ as
\begin{equation}
\rho \propto \frac{1}{a^{3(1+w)}}.
\end{equation}
 Since $w_r=1/3$ and $w_m=0$ for
radiation and non-relativistic matter, respectively, $\rho_r \propto 1/a^4$ while $\rho_m\propto 1/a^3$.
Therefore as the universe expands via growth of $a(t)$, at some point
\begin{equation}
\frac{\rho_m(t)}{\rho_m(t)+\rho_r(t)}
\end{equation}
trends to $1$ and energy density is almost entirely in matter. This occurs at the time of MRE, and a necessary
condition for this to occur is that the dark matter is actually behaving like matter at or prior to MRE time.
Since at MRE time $H(a_{MRE})\simeq 10^{-28}\, \text{eV}$, we have
\begin{equation}
m_\phi \geq 10^{-28}\, \text{eV}.
\end{equation}
An axion lighter than this scale could exist and potentially cause cosmological problems, but it cannot
be the dark matter that drives structure formation in our universe.

If the axion is all of the dark matter, or a non-trivial component of it, it must saturate or slightly
undersaturate the observed dark matter relic abundance \cite{Ade:2015xua}
\begin{equation}
\Omega_{cdm}h^2 = 0.12,
\end{equation}
where the Hubble constant today is $H_0=h \times 100 \text{km}/\text{s}/\text{Mpc}$ where $h\simeq .7$.
For an axion relic produced by the misalignment mechanism with $O(1)$ misalignment angle, its
abundance is
\begin{equation}
\Omega_{ax}h^2 = .7^2\times .1\times \left(\frac{F}{10^{17}\, \text{GeV}}\right)^2\, \left(\frac{m_\phi}{10^{-22}\, \text{eV}}\right)^\frac12.
\end{equation}
The experimental constraint is
\begin{equation}
\Omega_{ax} h^2 \leq \Omega_{cdm} h^2 = 0.12
\end{equation}
For the decay constants a few orders of magnitude below the Planck scale, as is the case for many string theoretic
axions, we see that the relevant axion mass scale is $m_\phi\simeq 10^{-22} \, \text{eV}.$ This mass scale is 
the relevant mass scale for so-called fuzzy dark matter, which is a form of ultralight dark matter that may 
account for a number of potential astrophysical anomalies on galactic scales. In \cite{Hui:2016ltb} it was shown that
string axions may give rise this scenario due to large instanton suppression factors; we refer the
reader to \cite{Hui:2016ltb} for a detailed account of the scenario and related astrophysical issues. 

One of the main results is that for decay constants typical in string theory, axions of the mass scale relevant for
fuzzy dark matter saturate the relic abundance. While perhaps not as strong of a coincidence as the so-called
WIMP miracle, it is certainly interesting. In \cite{Hui:2016ltb} the mass scale was accommodated by a very large instanton
suppression factor. In \cite{Halverson:2017deq} concrete SUSY models were presented and it was shown that stringy instantons that couple the axion to the
Higgs can trade very large instanton suppressions for smaller instanton suppressions and powers
of the electroweak hierarchy; this is the EFT analysis presented briefly above.

\subsection{Moduli}
\label{app:moduli}
In four-dimensional compactifications of string theory that preserve  $\mathcal{N}=1$ supersymmetry all scalars
must be complex, and therefore pseudoscalar axions must be
accompanied by scalar moduli. Varying the vacuum expectation value of these moduli fields corresponds to
varying the shape or size of certain cycles in the extra dimensions.  Moduli and axions together correspond to flat directions
in the leading order scalar potential, and therefore they must be stabilized by subleading effects such as fluxes
or instantons in order to avoid fifth force constraints. Like axions, the number of moduli fields is related
to topological numbers of the compactification manifolds $M$. Their existence is a generic prediction of string
compactifications.

If excited in the early universe, moduli can have important cosmological consequences. Here we briefly review moduli
in the case of Calabi-Yau threefolds and the cosmological consequences of these and other moduli. 
We refer the reader to \cite{Easther:2013nga,Kane:2015jia} for 
further details related to the discussion in this section.

\vspace{1cm}
Let $M$ be a compact Calabi-Yau threefold. These are six-manifolds with a Ricci-flat K\" ahler metric. In
the generic case they have holonomy $SU(3)$, though in some cases (for example, $M=K3\times T^2$) the holonomy
is $SU(2)$, and also a covariantly constant spinor. Actually finding such a metric on the compact space is
notoriously difficult, and currently there are no known non-trivial examples. Fortunately, Yau's theorem
guarantees the existence of such a metric, even if it cannot be computed. The theorem is that such a metric
exists if the first Chern class of the tangent bundle $TM$ is trivial, i.e., 
\begin{equation}
c_1(TM)=0.
\end{equation}
This topological condition is straightforward to check in many examples, for example if $M$ is a hypersurface 
or complete intersection in a higher-dimensional toric variety.
Compactifications of heterotic and Type $II$ string theory
on $M$ have $\cN=1$ and $\cN=2$ supersymmetry, respectively, and in Type $II$ compactifications the supersymmetry
may be broken to $\cN=1$ by the introduction of $1/2$-BPS branes and orientifolds. The choice of $M$ gives a
vacuum in four dimensions \cite{Candelas:1985en}.

From the point of view of the $10d$ theory, Calabi-Yau moduli correspond to infinitesimal 
deformations of the Ricci-flat
K\" ahler metric that lead to another Ricci-flat K\" ahler metric, and therefore to another Calabi-Yau
manifold and associated vacuum state. Under Kaluza-Klein reduction, metric deformations become $4d$ scalar
fields that are the moduli fields. There are two essential types of deformations: complex structure deformations
and K\" ahler deformations. These give rise to complex structure and K\" ahler moduli whose number are
\begin{align}
\text{\# Complex Structure Moduli}&= h^{21}(M) \nonumber \\
\text{\# K\" ahler Moduli}&= h^{11}(M),
\end{align}
where $h^{21}(M)$ and $h^{11}(M)$ are the Hodge numbers of the Calabi-Yau threefold; see, e.g., \cite{griffiths2011principles}. These typically number in the dozens or hundreds, as discussed. The largest known database of Calabi-Yau
threefolds are those corresponding to hypersurfaces in Gorenstein Fano toric varieties. The number is not
known, but the Gorenstein Fano toric varieties can be associate to fine regular star triangulations
of $4d$ reflexive polytopes. The latter have been classified by Kreuzer-Skarke \cite{Kreuzer:2000xy}, and there are $473,800,776$ of them.

\vspace{1cm}

From the point of view of $4d$ cosmology, the critical difference between the modulus and axion field that
belong to the same complex scalar is that the axion enjoys a shift symmetry while the modulus does not. The axion
therefore receives its mass from non-perturbative effects (instantons) that break the perturbative shift symmetry,
whereas many effects may contribute to the modulus mass. For this reason, moduli are usually heavier than axions,
and in many models the lightest modulus $\phi$ satisfies
\begin{equation}
m_\phi \simeq m_{3/2},
\end{equation}
where $m_{3/2}$ is the gravitino mass. This relationship is typical of models with gravity mediated supersymmetry
breaking. In some models, which are often referred to as sequestered models \cite{Kachru:2007xp}, the relationship may be broken,
but sequestering is difficult to achieve in string compactifications \cite{Berg:2010ha}.

A central aspect of cosmologies with moduli is the so-called cosmological moduli problem \cite{Coughlan:1983ci,deCarlos:1993wie,Banks:1993en}. Similar to the case of
an oscillating axion, when the modulus oscillates about its minimum it behaves like cold matter  and will
therefore come to dominate the energy density of the universe, due again to the $\rho_\phi \propto 1/a^3$ scaling.
If the eventual decay of the modulus does not reheat the universe to a temperature above the BBN temperature
$T_{BBN}\simeq 5 \, \text{MeV}$ then the successful predictions of nucleosynthesis are spoiled. This 
is problem is known as the cosmological moduli problem. One solution is to eliminate moduli from the theory,
but this is not well-motivated from the perspective of string theory.

The simplest solution to the cosmological moduli problem in a string-motivated scenario is for the
modulus to be massive enough that the reheat temperature associated with its decay is above
the BBN temperature. Since moduli
couple to the visible sector via non-renormalizable operators we will study decays that use a dimension-$5$
operator. The rate is
\begin{equation}
\Gamma_\phi = c_\phi^2 \frac{m_\phi^3}{M_{pl}^2},
\end{equation}
where $c_\phi$ is the Wilson coefficient of the effective operator.
If the $\phi$ decay and subsequent thermalization are instantaneous
then the reheat temperature is
\begin{equation}
T_{rh} = \left[ \left(\frac{8}{90} \pi^3 g_\star\right)^{-1/2} \, M_{pl} \, \Gamma_\phi\right]^{1/2} 
\simeq 10 \, c_\phi \, \left(\frac{m_\phi}{100\, \text{TeV}}\right)^{3/2}\, \text{MeV}.
\end{equation}
The nucleosynthesis bound $T_{rh}\geq T_{BBN} \simeq 5\, \text{MeV}$ gives
\begin{equation}
m_\phi \gtrsim 100\, \left(\frac{1}{2c_\phi}\right)^{2/3} \, \text{TeV} = 63\, \text{TeV},
\end{equation}
where the last equality is under the assumption $c_\phi = 1$. The bound can be moved slightly
due to factor of $2$ uncertainties in $T_{BBN}$ or changes in the Wilson coefficient or assumption
of instantaneous thermalization. However, the qualitative conclusion that
\begin{equation}
m_\phi \gtrsim \, O(50) \, \text{TeV}
\label{eqn:modbound}
\end{equation}
does not depend strongly on such modifications.

Even if the cosmological moduli problem is avoided by sufficiently heavy modulus,
there are at least two significant ways in which particle physics and cosmology are affected
by the modulus.

Consider a theory where SUSY solves the electroweak hierarchy problem
that also has a modulus satisfying the bound \eqref{eqn:modbound}. If the model gives rise to
the relation
\begin{equation}m_\phi \simeq m_{3/2}\end{equation} 
then the bound implies that the supersymmetry breaking scale is an order of magnitude above the
TeV scale. Such scenarios have been extensively studied and typically exhibit heavy scalar 
partners of the standard model fermions \cite{Wells:2003tf,ArkaniHamed:2004fb,Langacker:2007ac,Kane:2011kj}.

The cosmological implications of the heavy modulus do not require weak-scale SUSY, but are compatible with
it. In a standard cosmological history the universe is radiation dominated during nucleosynthesis and the first
phase of matter domination occurs at matter-radiation equality, as discussed. However, it is also possible
that there is a phase of matter domination \emph{prior} to nucleosynthesis, in which case other aspects of
cosmology may also be affected. For example, WIMP dark matter produced by thermal freezeout has a freezeout
temperature
\begin{equation}
T_{fo}\simeq \frac{m_\chi}{20},
\end{equation}
where $\chi$ is the WIMP. If the relation
\begin{equation}
T_{BBN} \leq T_{rh} \leq T_{fo}
\label{eqn:nonthermal}
\end{equation}
is satisfied then the decaying modulus avoids the cosmological moduli problem, but reheats the universe to 
a temperature below the scale necessary to produce $\chi$ via by thermal freeze-out. Instead, 
the dark matter $\chi$ may
be produced directly by modulus decay, in which case the dark matter phenomenology changes, in particular
predictions for indirect detection. If $m_\chi$ is electroweak scale, as is typical for WIMPs, the window
\eqref{eqn:nonthermal} is three orders of magnitude.

\subsection{Exotics and Consistency Conditions}
\label{app:consistency}
In this section we discuss certain string consistency
conditions, their relation to consistency conditions
in field theory, and their implications for particle
spectra. We will begin with a concrete study of consistency
conditions and will then proceed to understand them multiple
points of view. We will then discuss implications of the
conditions for embeddings of the minimal supersymmetric
standard model (MSSM) into string theory.We will discuss multiple duality frames, but since
the derivation relevant to us is straightforward in
Type IIA compactifications, we begin there.

\subsubsection*{String Consistency Conditions}

Consider $4d$ compactifications of the Type IIA superstring
on a Calabi-Yau threefold $M$ with O6-planes and D6-branes.
The orientifold requires an antiholomorphic involution
\begin{equation}
\sigma: M \to M
\end{equation}
such that the fixed locus of $\sigma$ is a three-cycle
\begin{equation}
\sigma_{\text{fixed}}=: \pi_{O6} \qquad [\pi_{O6}] \in H_3(M,\bZ).
\end{equation}
The spacetime-filling O6-plane on $\pi_{O6}$ is magnetically (electrically) charged under the
the Ramond-Ramond one-form (seven-form) $C_1$ ($C_7$).
Since $M$ is compact, Gauss' law requires that there are
other charged objects on $M$ so that the net $C_7$-charge
is zero. D6-branes may be added to the compactification for
this purpose, and the Gauss law condition --- also known
as D6-brane tadpole cancellation --- is
\begin{equation}
\sum_a N_a \left(\pi_a + \pi_a'\right) = 4\pi_{O6},
\label{eqn:tadpole}
\end{equation}
where there are $N_a$ D6-branes on the three-cycle $\pi_a$
and also the orientifold image cycle $\pi_a'$. Finally, if each $\pi_a$
is a special Lagrangian submanifold and an additional condition
is satisfied, presence of the D6-branes and O6-plane breaks
the $\cN=2$ supersymmetry of the Calabi-Yau background to
$\cN=1$.

\begin{table}
\centering \vspace{3mm}
\label{table:spectrum}
\begin{tabular}{|c|c|}
\hline
Representation  & Multiplicity \\
\hline $\Yasymm_a$
 & ${1\over 2}\left(\pi_a\circ \pi'_a+ \pi_a\circ \pi_{{\rm O}6}
\right)$  \\
 $\Ysymm_a$
      & ${1\over 2}\left(\pi_a\circ \pi'_a-\pi_a\circ \pi_{{\rm O}6}
\right)$   \\
      $(\fund_a,\antifund_b)$
       & $\pi_a\circ \pi_{b}$   \\
        $(\fund_a,\fund_b)$
	 & $\pi_a\circ \pi'_{b}$
	 \\
	 \hline
	 \end{tabular}
	 \caption{Representations and multiplicities for chiral matter at the intersection of two D6-branes.}
	 \label{table:intersections}
\end{table}

For general $\pi_a$ the stack of $N_a$ D6-branes carries a $U(N_a)$ gauge group, though it may carry a symplectic or special orthogonal group if $\pi_a$ is orientifold invariant. We consider the case of unitary groups, and therefore
\begin{equation}
G = \prod_a U(N_a).
\end{equation}
The trace $U(1)_a\subset U(N_a)$ is often anomalous, in which
case it receives a mass via the St\" uckelberg mechanism.
However, under certain conditions a linear combination
\begin{equation}
U(1)_q = \sum_a q_a U(1)_a
\end{equation}
is left massless; see, e.g., \cite{Cvetic:2011vz} for additional
details.
Charged chiral superfields may arise at the intersection of
D6-brane stacks or at the intersection of D6-branes with
the O6-plane. The possible representations are bifundamentals
of $U(N_a)\times U(N_b)$ or two-index symmetric or
antisymmetric tensor representations of $U(N_a)$.
The chiral index of each such representation are 
counted by intersection numbers in $M$, which are presented
in Table \ref{table:intersections}.

We may now study the relationship between string consistency
conditions and constraints on chiral spectra. The Gauss
law (D6-brane tadpole cancellation) \eqref{eqn:tadpole}
is necessary for consistency of the string compactification,
and it puts a single constraint on 
the homology cycles wrapped by the D6-branes and O6-plane.
This equation that constraints homology may be turned into
a number of equations that constrain integers representing
the chiral spectra by intersecting \eqref{eqn:tadpole}
with $\pi_b$, in which case a short computation using
Table \ref{table:spectrum} yields
\begin{equation}
	\label{eqn:chiral tadpole constraint}
	T_b := \# b - \# \ov b + (N_b+4)\,\, (\# \, \Ysymm_b - \#\, \ov \Ysymm_b) + (N_b-4) \,\, (\# \, \Yasymm_b - \# \, \ov \Yasymm_b) = 0.
\end{equation}
Since this equation follows from \eqref{eqn:tadpole} is is
necessary for D6-brane tadpole cancellation. It constrains
the chiral spectrum of the $4d$ $U(N_b)$ gauge theory 
arising from $N_b$ D6-branes on $\pi_b$. 

\vspace{1cm}
Consistency constraints on chiral spectra are commonplace
in quantum field theory, usually representing an anomaly
cancellation condition.  Before attempting to understand
\eqref{eqn:chiral tadpole constraint} from the point
of view of anomalies, we would first like to recall
some properties of anomalies for $SU(N)$ gauge theories
in four dimensions. Consider left-handed Weyl fermions
in a representation $R$ of $SU(N)$ and let 
\begin{equation}\chi(R)=\# R - \# \ov R\end{equation}  
be its chiral index. Then the $SU(N)^3$ anomaly cancellation
takes the form
\begin{equation}
\sum_R A_R \, \chi(R) = 0,
\label{eqn:anom}
\end{equation}
where $A_R$ is the anomaly coefficient associated to the
representation $R$. For the fundamental, $\Ysymm$,
and $\Yasymm$ representations it is $1$, $(N+4)$, and $(N-4)$,
respectively. 

However, for this constraint to be non-trivial,
complex conjugate representations $\ov R$ must exist, so
that $\chi(R)$ can be non-zero. For $N>2$, such representations
exist and the constraint is non-trivial. $SU(2)$, on the
other hand, only has self-conjugate representations,
and accordingly there is no $SU(2)^3$ anomaly or associated
cancellation condition. Furthermore, $SU(2)$ gauge theories
may have so-called Witten anomalies \cite{Witten:1982fp}, which arises 
from the fact that
\begin{equation}
\pi_4(SU(2)) = \bZ_2,
\end{equation}
and in particular that it is non-zero. For example,
if there are an odd number of Weyl fermion doublets
of $SU(2)$ the large gauge transformation associated with
$\pi_4(SU(2))\neq 0$ introduces a sign into the path
integral, forcing it to be zero. $SU(N>2)$ gauge
theories in four dimensions cannot have such anomalies,
since
\begin{equation}
\pi_4(SU(N>2))=0.
\end{equation}
In summary, cubic non-abelian anomalies of type $SU(N)^3$ may
only exist for $N>2$, whereas Witten anomalies may only
exist for $N=2$. Anomaly cancellation conditions in $4d$
$SU(N)$ gauge theories are fundamentally different according
to whether $N=2$ or $N>2$.

Since $U(N_a)$ gauge theories arising from $N_a$ D6-branes
in weakly coupled Type IIA compactifications may only have (chiral) 
fundamental
representations (that arise as a bifundamental with another group) or two
index symmetric or antisymmetric tensor representations, only those
representations should appear in an anomaly cancellation condition
for the $SU(N_a)$ subgroup of $U(N_a)$. This is precisely \eqref{eqn:chiral tadpole constraint}
in the case $N_a>2$. The constraint $T_b=0$, which is necessary for D6-brane tadpole
cancellation, ensures the absence of $SU(N_b)^3$ anomalies in the case $N_b>2$.

On the other hand, the constraints for $U(2)$ and $U(1)$, which we call $T_2=0$ and $T_1=0$, respectively,
are less obvious from a field theoretic point of view.
As is often said, tadpole cancellation is stronger than anomaly cancellation
\cite{Uranga:2000xp,Aldazabal:2000dg}, and these constraints $T_2=T_1=0$ are often
referred to as ``stringy'' constraints. These stringy constraints 
have been studied in a number of works, e.g.,  \cite{Aldazabal:2000dg,Uranga:2000xp,Anastasopoulos:2006da,Cvetic:2011iq,Halverson:2013ska}. 
A summary of the current understanding of the stringy constraints
is that they are:
\begin{itemize}
\item Necessary but not sufficient for tadpole
cancellation. 
\item Necessary but not sufficient for $U(1)_a\subset U(N_a)$
anomaly cancellation. This constraint on the chiral spectrum cancels the part of the
anomaly not canceled by Chern-Simons terms.
\item Necessary and sufficient
for $SU(M+N_a)^3$ anomaly cancellation if the
system were to nucleate $M$ brane anti-brane pairs, embedding
$U(N_a)$ into $U(M+N_a)$. 
\end{itemize}
Each of these pathologies would render the theory inconsistent if the conditions were violated.

\vspace{1cm}
Before proceeding to a discussion of phenomenological implications of these constraints, we would like to make
two additional comments.

First, similar constraints also arise in other corners of the landscape. For example, in Type IIB 
orientifold compactifications with magnetized D7-branes and D3-branes, the D7- and D5-tadpole cancellation
conditions together play the same role that the D6-brane tadpole cancellation condition plays in Type
IIA compactifications. From those conditions, \eqref{eqn:chiral tadpole constraint} can be derived in
a different duality frame. In heterotic compactifications similar constraints are also known to exist 
\cite{Blumenhagen:2005pm}. From the Type IIA and Type IIB constraints it is also likely that they can
be lifted to strong coupling, that is to M-theory and F-theory, though the form of \eqref{eqn:chiral tadpole constraint}
may be more general due to the possible addition of more representations to the spectrum at strong coupling.

The constraints can be conceptually understood using one of the deep insights from the second string
revolution: gauge theories that arise in string compactifications are often carried by charged objects. If these
charged objects pair produce and change the structure of the gauge theory, there may be additional constraints
necessary for consistency of the gauge theory after pair production. For example, if an $SU(2)$ gauge theory is 
the stable end point of a process by which an $SU(N)$ theory on $N+2$ D-branes collides with an $SU(2)$ theory
on $2$ anti-branes, the $SU(N)$ theory prior to the collision must also be anomaly free. This can impose
constraints on the $SU(2)$ theory that would not be present from a purely bottom up perspective.
See \cite{Schwarz:2001sf,Uranga:2000xp,Halverson:2013ska} for additional calculations and discussions that utilize this idea.

Finally, string constraints together with supersymmetry conditions can lead to interesting
finiteness results. For example, in type IIA compactifications on a toroidal orbifold the
D6-brane tadpole cancellation conditions together with the supersymmetry conditions yield
a finite number of possible gauge sectors \cite{Douglas:2006xy}. Similar results  also
apply in type IIB compactifications with magnetized D9-branes on a number of smooth Calabi-Yau
threefold elliptic fibrations \cite{Cvetic:2014gia}. It is likely that such results hold
quite generally in the landscape as a result of consistency conditions and supersymmetry
conditions, which is an interesting direction for future work.

\subsubsection*{Phenomenological Implications}

The conditions \eqref{eqn:chiral tadpole constraint} constraint the chiral spectrum of $U(N_a)$ gauge theories
realized in the landscape, and as we have discussed the constraints $T_2=0$ and $T_1=0$ on $U(2)$ and $U(1)$
gauge theories are ``stringy,'' i.e., they do not correspond to any obvious or typical constraint in quantum
field theory.

In this section we study the implications of the stringy constraints for possible embeddings of the MSSM
into string theory. We will aim for brevity and clarity, since the details of this section have been
discussed elsewhere \cite{Cvetic:2011iq}. 

Consider a bottom-up embedding of an MSSM-like model, where stacks of D-branes are engineered such that the visible sector gauge group
is
\begin{equation}
G = U(3)_a \times U(2)_b \times \prod_{i=1}^{N} U(1)_i,
\end{equation}
where $SU(3)_{QCD}\subset U(3)_a$ and $SU(2)_L\subset U(2)_b$. These are $(N+2)$-stack models. 
The weak hypercharge arises as a linear combination
\begin{equation}
Y = y_a U(1)_a + y_b U(1)_b + \sum_{i=1}^N y_i U(1)_i,
\end{equation}
and for models with a low number of stacks the possible hypercharge linear combinations, which
are referred to as hypercharge embeddings, have been classified
\cite{Anastasopoulos:2006da}. There are additional conditions that must be satisfied \cite{Aldazabal:2000dg} for this
linear combination to remain massless.

Given a fixed $N$ and hypercharge embedding, together with the fact that chiral superfields may arise as 
bifundamental or two-index symmetric or antisymmetric tensor representations, possible realizations of each MSSM
chiral superfield may be classified. From that classification, the possible embeddings of the MSSM into the
string construction may be classified, and by direct calculation one finds that nearly all
violate the $T_2=0$ or $T_1=0$ conditions.

In a number of works, including works by the authors, only models
where the MSSM superfields satisfy the $T_2=0$ and $T_1=0$ conditions were studied.
However, models that violate these stringy constraints are not ruled out and the associated MSSM spectra may arise
as spectra in consistent string compactifications. The caveat is that they must be accompanied by other fields charged
under $G$ that also contribute to the $T_2$ or $T_1$ condition, so that
\begin{equation}
T_2 = T_{2,\text{MSSM}}+T_{2,\text{exotics}}=0
\qquad \qquad T_1 = T_{1,\text{MSSM}}+T_{1,\text{exotics}}=0,
\end{equation}
even if $T_{2,\text{MSSM}}\neq0$ or $T_{1,\text{MSSM}}\neq 0$. In these models, the string constraints are 
requiring the existence of $G$-charged exotics, which usually are also charged under 
$G_{SM}=SU(3)\times SU(2)\times U(1)$.

In \cite{Cvetic:2011iq}, a systematic study was done of all consistent quivers of 
this type that contain a three family MSSM spectrum with a single
pair of Higgs doublets and up to five $G$-charged chiral superfields beyond the
MSSM. There are $89964$ such models, and nearly all have exotics beyond the MSSM
that are required so that $T_{2,\text{MSSM}}\neq0$ or $T_{1,\text{MSSM}}\neq 0$ is canceled;
i.e., these exotics are added for string consistency.

The results are presented in Table \ref{table:particle addition table}.
Relevant comments on the results include:
\begin{itemize}
\item String constraints have clear preferences for some exotic multiplets over others.
\item Chiral fourth families and hyper-charge shifted fourth families occur relatively infrequently.
\item The most common exotic is a standard model singlet, and it occurs far more than any other
	exotic. Though they are a singlet with respect to the standard model, they are always
	charged under anomalous $U(1)$'s, which dictate their couplings to other MSSM fields.
\item Next most frequent is a colorless $SU(2)_L$ triplet with $Y=0$.
\item From there, we a see a variety of exotic quarks and leptons, and from their multiplicities
	we infer that they are coming in vector-like pairs with respect to the standard model.
	However, by the assumptions of the analysis they are always chiral with respect to some anomalous $U(1)$.
\end{itemize}

In summary, we have seen that string consistency conditions can place constraints on chiral spectra that
force the existence of additional exotics beyond the MSSM particles. They are often quasi-chiral,
which are of phenomenological interest, as discussed in the main text.

\begin{table}[htb]
\centering
\scalebox{.95}{
\begin{tabular}{|c|c|c|c|c|}\hline
SM Rep & Total Multiplicity & Int. El. & $4^\text{th}$ Gen. Removed & Shifted $4^\text{th}$ Gen. Also Removed \\ \hline \hline
$(\textbf{1},\textbf{1})_{0}$ & $174276$ & $173578$ & $173578$ & $173578$ \\ \hline
$(\textbf{1},\textbf{3})_{0}$ & $48291$ & $48083$ & $48083$ & $48083$ \\ \hline
$(\textbf{1},\textbf{2})_{-\frac{1}{2}}$ & $39600$ & $39560$ & $38814$ & $38814$ \\ \hline
$(\textbf{1},\textbf{2})_{\frac{1}{2}}$ & $38854$ & $38814$ & $38814$ & $38814$ \\ \hline
$(\ov{\textbf{3}},\textbf{1})_{\frac{1}{3}}$ & $25029$ & $25007$ & $24261$ & $24241$ \\ \hline
$(\textbf{3},\textbf{1})_{-\frac{1}{3}}$ & $24299$ & $24277$ & $24277$ & $24241$ \\ \hline
$(\textbf{1},\textbf{1})_{1}$ & $15232$ & $15228$ & $14482$ & $14482$ \\ \hline
$(\textbf{1},\textbf{1})_{-1}$ & $14486$ & $14482$ & $14482$ & $14482$ \\ \hline
$(\ov{\textbf{3}},\textbf{1})_{-\frac{2}{3}}$ & $3501$ & $3501$ & $2755$ & $2755$ \\ \hline
$(\textbf{3},\textbf{1})_{\frac{2}{3}}$ & $2755$ & $2755$ & $2755$ & $2755$ \\ \hline
$(\textbf{3},\textbf{2})_{\frac{1}{6}}$ & $1784$ & $1784$ & $1038$ & $1038$ \\ \hline
$(\ov{\textbf{3}},\textbf{2})_{-\frac{1}{6}}$ & $1038$ & $1038$ & $1038$ & $1038$ \\ \hline
$(\textbf{1},\textbf{2})_{0}$ & $852$ & $0$ & $0$ & $0$ \\ \hline
$(\textbf{1},\textbf{2})_{\frac{3}{2}}$ & $220$ & $220$ & $220$ & $184$ \\ \hline
$(\textbf{1},\textbf{2})_{-\frac{3}{2}}$ & $204$ & $204$ & $204$ & $184$ \\ \hline
$(\textbf{1},\textbf{1})_{\frac{1}{2}}$ & $152$ & $0$ & $0$ & $0$ \\ \hline
$(\textbf{1},\textbf{1})_{-\frac{1}{2}}$ & $152$ & $0$ & $0$ & $0$ \\ \hline
$(\textbf{3},\textbf{1})_{\frac{1}{6}}$ & $124$ & $0$ & $0$ & $0$ \\ \hline
$(\ov{\textbf{3}},\textbf{1})_{-\frac{1}{6}}$ & $124$ & $0$ & $0$ & $0$ \\ \hline
$(\textbf{3},\textbf{1})_{-\frac{4}{3}}$ & $36$ & $36$ & $36$ & $0$ \\ \hline
$(\textbf{1},\textbf{3})_{-1}$ & $36$ & $36$ & $36$ & $0$ \\ \hline
$(\ov{\textbf{3}},\textbf{2})_{\frac{5}{6}}$ & $36$ & $36$ & $36$ & $0$ \\ \hline
$(\ov{\textbf{3}},\textbf{1})_{\frac{4}{3}}$ & $20$ & $20$ & $20$ & $0$ \\ \hline
$(\textbf{1},\textbf{3})_{1}$ & $20$ & $20$ & $20$ & $0$ \\ \hline
$(\textbf{3},\textbf{2})_{-\frac{5}{6}}$ & $20$ & $20$ & $20$ & $0$ \\ \hline
\end{tabular}}
\caption{Displayed are the standard model representation of matter additions obtained in \cite{Cvetic:2011iq}, together with their total multiplicity
  across all three-node and four-node quivers. The third column excludes quivers
  involving states that would lead to fractionally-charged color singlets, which are strongly constrained by
  cosmology. The fourth further excludes those where the matter additions correspond to a chiral fourth generation, while the last also excludes a hypercharge-shifted fourth generation. The additions that remain after
  imposing these cuts make up most of the set, and are comprised of MSSM singlets, $SU(2)$ triplets with $Y=0$, and  quasichiral pairs.}\label{table:particle addition table}
\end{table}

\bibliographystyle{utphys} 
\bibliography{Remnants}
\end{document}